\documentclass[twocolumn,superscriptaddress,showkeywords,preprintnumbers,amsmath,amssymb,aps,pre]{revtex4}
\usepackage{graphicx}
\usepackage{dcolumn}
\usepackage{bm}
\begin{document}
\title{Investigation of the seismicity after the initiation of a  Seismic Electric Signal activity until the main shock}
\author{N. V. Sarlis}
\affiliation{Solid State Section and Solid Earth Physics Institute, Physics Department, University of Athens, Panepistimiopolis, Zografos 157 84,
Athens, Greece}
\author{E. S. Skordas}
\affiliation{Solid State Section and Solid Earth Physics Institute, Physics Department, University of Athens, Panepistimiopolis, Zografos 157 84,
Athens, Greece}
\author{M. S. Lazaridou}
\affiliation{Solid State Section and Solid Earth Physics Institute, Physics Department, University of Athens, Panepistimiopolis, Zografos 157 84,
Athens, Greece}
\author{P. A. Varotsos}
\thanks{{\bf Correspondence to:} P. Varotsos (pvaro@otenet.gr)}
\affiliation{Solid State Section and Solid Earth Physics Institute, Physics Department, University of Athens, Panepistimiopolis, Zografos 157 84,
Athens, Greece}

\begin{abstract}
The behavior of seismicity in the area candidate to suffer a main shock is investigated after the observation of the Seismic Electric Signal activity
until the impending mainshock. This makes use of the concept of natural time $\chi$ and reveals that the probability density function 
of the variance $\kappa_1(=\langle \chi^2 \rangle -\langle \chi \rangle
^2)$ exhibits distinct features before the occurrence of the mainshock. Examples are presented, which refer 
to magnitude class 6.0 earthquakes that occurred in Greece during the first two months in 2008.

\end{abstract}                                                           
\maketitle

\section{Introduction}
Seismic Electric Signals (SES) are
transient low frequency ($\leq$ 1Hz) electric signals that have been observed in Greece\cite{proto,proto2,var88x,VAR91,VAR93,var99,grl}, Japan\cite{uye,uye2}, Mexico\cite{MEX07}  etc. days to months  before
earthquakes(EQs).  They are
emitted when the stress in the focal region reaches a {\em critical}
value before the failure\cite{varbook,newbook}. This stems from the fact that a 
stress variation affects the Gibbs energy for the defect formation\cite{VAR77}, migration\cite{VAR78} and activation\cite{VAR85} in solids. For 
EQs with magnitude 6.5 or larger, SES are accompanied by detectable magnetic field variations\cite{PJA1,PJA2,PRL03}.  
A sequence of SES observed within a short time (e.g. $\approx 1$h) is termed SES activity the analysis of which has been shown to obey
an $1/f$-behavior\cite{NAT02,WER05}. 
Recently, a method\cite{NAT01,NAT02A,VAR05C,NAT06A,NAT06B,JAP08}  has been presented that 
enables the shortening of the time-window of the impending mainshock from a few hours to a few days only. 
It is based on the concept of a new time domain, termed natural time\cite{NAT01,NAT02,NAT02A} and investigates the order parameter of seismicity\cite{VAR05C} (see also below) that occurs after the SES activity and before the main shock in the area candidate to suffer a strong  EQ. The improvement of that method constitutes the basic aim of the present paper in view of the great practical importance 
in determining the time of an impending catastrophe. Along these lines, 
the most recent SES electric field data are also presented.

In a time series consisting of $N$ events, the {\em natural time}
$\chi_k = k/N$ serves as an index\cite{NAT01,NAT02,NAT02A} for the
occurrence of the $k$-th event. The evolution of the pair
($\chi_k, Q_k$) is studied\cite{NAT01,NAT02,NAT02A,NAT03,NAT03B,NAT04,NAT05,NAT05B,VAR05C,NAT06A,NAT06B,JAP08,newbook},
where $Q_k$ denotes a quantity proportional to the energy released
in the $k$-th event. For dichotomous signals, for example, which
is frequently the case of SES activities, $Q_k$ can be replaced by the
duration of the $k$-th pulse. As a second example, we refer to the analysis of seismicity\cite{NAT01,NAT02A,varbook,TAN04},  where 
$Q_k$ may be considered as the seismic moment $M_{0 k}$
of the 
$k$-th event, since  $M_{0}$ is roughly proportional to the energy released during an EQ.
The normalized power spectrum is given\cite{NAT01,NAT02,NAT02A} by $\Pi(\omega )\equiv | \Phi (\omega ) |^2 $, where
$\Phi (\omega )$ is defined as 
\begin{equation}
\label{eq3} \Phi (\omega)=\sum_{k=1}^{N} p_k \exp \left( i \omega
\frac{k}{N} \right)
\end{equation}
In this definition, $p_k$ stands for  $p_k=Q_{k}/\sum_{n=1}^{N}Q_{n}$, and $\omega =2 \pi \phi$; where $\phi$
denotes the {\it natural frequency}. The continuous function
$\Phi (\omega )$ in Eq.(\ref{eq3}) should {\em not} be confused with the usual
discrete Fourier transform because the latter considers only the relevant values at
$\phi=0,1,2,\ldots$, while  in natural time analysis the properties of
$\Pi(\omega)$ or $\Pi(\phi)$ are studied\cite{NAT01,NAT02,NAT02A,newbook}  for
natural frequencies $\phi$ less than 0.5. This is so, because in
this range of $\phi$, $\Pi(\omega)$  or $\Pi(\phi)$ reduces
 to a {\em characteristic function} for the
probability distribution $p_k$  in the context of probability
theory.

 When the system enters the
{\em critical} stage, the following relation
holds\cite{NAT01,NAT02,VAR05C}:
\begin{equation}
\Pi ( \omega ) = \frac{18}{5 \omega^2} -\frac{6 \cos \omega}{5
\omega^2} -\frac{12 \sin \omega}{5 \omega^3}, \label{fasma}
\end{equation}
which for $\omega \rightarrow 0$, simplifies to\cite{NAT01,NAT02,newbook}
\[ \Pi (\omega )\approx 1-0.07
\omega^2.\]
This relation  reflects\cite{VAR05C} that the variance $\langle \chi^2 \rangle -\langle \chi \rangle
^2$ of $\chi$
is given by
\begin{equation}
 \kappa_1=\langle \chi^2 \rangle -\langle \chi \rangle
^2=0.07, \label{ex3}
\end{equation}
 where $\langle f( \chi) \rangle = \sum_{k=1}^N p_k
f(\chi_k )$. Note that in the case of seismicity, Eq.(\ref{fasma}) was found\cite{VAR05C} to describe adequately the most 
probable value of $\Pi(\omega)$. Furthermore, as shown in Ref.\cite{VAR05C},  $\Pi(\omega)$ for $\omega \rightarrow 0$ (or $\kappa_1$) can be considered as an order
parameter for seismicity since its value changes abruptly when a main shock occurs and the statistical properties 
of its fluctuations resemble those in other nonequilibrium systems (e.g., three-dimensional
turbulent flow) as well as in  equilibrium critical
phenomena (e.g., two-dimensional Ising model).

Apart from $\Pi(\omega)$ or $\kappa_1$, another useful quantity in natural time is the entropy $S$, which  is defined
as\cite{NAT01,NAT03B} \[ S \equiv  \langle \chi \ln \chi \rangle -
\langle \chi \rangle \ln \langle \chi \rangle.\] 
This quantity depends on
the sequential order of events\cite{NAT04,NAT05} and exhibits\cite{NAT05B} concavity, positivity and Lesche\cite{LES82,LES04}
stability. The $S$ value becomes equal to $\ln 2
/2-1/4\approx 0.0966$ for a ``uniform'' (u) distribution,  as it was defined
in Refs. \cite{NAT01,NAT03,NAT03B,NAT04,NAT05}, e.g.   when all
$p_k$ are equal or $Q_k$ are positive independent and identically
distributed random variables of finite variance (In this case, $\kappa_1$ and $S$ are designated $\kappa_u(=1/12)$ and $S_u(=\ln 2
/2-1/4)$, respectively). 
 The same holds for the value of the entropy obtained\cite{NAT05B,NAT06A}
upon considering the time reversal ${\mathcal T}$, i.e.,
${\mathcal T} p_k=p_{N-k+1}$, which is labelled by $S_-$. 
The SES activities, when analyzed in natural time, have been found to exhibit
{\em infinitely} ranged temporal correlations and -beyond Eq.(\ref{ex3})- obey the
conditions\cite{NAT06A,NAT06B}:
\begin{equation}\label{eq2}
    S, S_- < S_u.
\end{equation}

The present paper is organized as follows:
Section \ref{dyo} presents a new procedure for the study of the order parameter of seismicity($\kappa_1$) in the area 
candidate to suffer a main shock during the period after the initiation of the SES activity until the mainshock occurrence.
As examples, we apply this procedure in Section \ref{tria} to the most recent SES data. Finally, 
Section \ref{tesera} presents the conclusions.

\section{New Method for shortening the time-window by studying the order parameter of seismicity in the area candidate to suffer a main shock}
\label{dyo}

Earthquakes exhibit complex correlations in space, time and magnitude (M), as shown by many studies, e.g. \cite{VAR05C,NAT06A,NAT06B,JAP08,NAT03,NAT03B,NAT04,NAT05,NAT05B,TAN04,BAK02,COR04,BAI04,ABE04,SHE06,HOL06,TIA07}.
Moreover, it has been repeatedly proposed that the occurrence of earthquakes  can be considered as 
a critical point (e.g. \cite{KLE07}, see also Ref.\cite{SOR04} and references therein).
Natural time  reveals, as mentioned in Section I, that $\Pi(\omega)$ or $\kappa_1$ can be considered as an order
parameter  for seismicity\cite{VAR05C}.
To obtain $\Pi(\omega)$ or $\kappa_1$, however,    it is necessary to decide the initiation time of seismicity analysis. We decided to start the analysis immediately after the SES initiation. This is based on our fundamental premise proposed long ago\cite{proto,varbook,newbook} that both SES emission and earthquake occurrence are critical phenomenon and, in a sense, the approach to ``electrical'' critical point shortly precedes ``mechanical'' critical point. 

\subsection{Background}\label{set}
Once the SES activity has been recorded, an estimation of the area to suffer a mainshock can be obtained on the basis of the so-called selectivity 
map\cite{VAR91,DOL99,newbook,TEC05} of the station at which the SES obsrvation was made. Thus, we have some area, hereafter labelled A, in which we count the small events (earthquakes) $e_i$ that occur 
after the initiation of the SES activity. Each event $e_i$ is characterized by its location ${\bf x}(e_i)$, the conventional time  of its
occurrence $t(e_i)$, and its 
magnitude ${\rm M}(e_i)$ or the equivalent seismic moment $M_0(e_i)$. 
The index $i=1,2,\ldots, N$ increases by one each time a new earthquake  with magnitude $M$ larger or equal to some threshold $M_{thres}$ occurs within the
area A (cf. $\forall i, t(e_{i+1})> t(e_i)$). Thus, a set of events $\cal{A}$ is formed until the mainshock occurs in A at $i=N$.
 To be more precise, a family of
sets ${\cal A}_{{\rm M}_{thres}}$ should have been formed before the mainshock 
occurs in A at $i=N$, where ${\cal A}_{{\rm M}_{thres}}=\left\{ e_i \in {\cal A}: {\rm M}(e_i)\geq {\rm M}_{thres}\right\}$ 
and the number of events in ${\cal A}_{{\rm M}_{thres}}$ is denoted by $|{\cal A}_{{\rm M}_{thres}}|$. The set ${\cal A}_{{\rm M}_{thres}}$ becomes a (time) ordered set (i.e., the events are written in the sequence of their occurrence time) by selecting appropriately the indices $j$  $j$ for its elements $e_j$, $j=1,2,\ldots ,|{\cal A}_{{\rm M}_{thres}}|$  so that $\forall j, t(e_{j+1})> t(e_j)$.  
Since earthquakes do not occur everywhere within the area A but in some specific locations, we can also 
define $R({\cal A}_{{\rm M}_{thres}})$ as the minimal rectangular (in latitude and longitude) region in which the epicenters of  the events 
of ${\cal A}_{{\rm M}_{thres}}$ are located. Moreover, for a given ordered set ${\cal A}_{{\rm M}_{thres}}$ the corresponding values 
of $\kappa_1({\cal A}_{{\rm M}_{thres}})$, $S({\cal A}_{{\rm M}_{thres}})$ and $S_-({\cal A}_{{\rm M}_{thres}})$ can be obtained by analyzing 
in natural time its ordered elements $e_j$ ($\in {\cal A}_{{\rm M}_{thres}}$). This is made by analyzing in natural time the pairs  $(\chi_j,Q_j)=(j/|{\cal A}_{{\rm M}_{thres}}|, M_0(e_j))$ where $j=1,2,\ldots ,|{\cal A}_{{\rm M}_{thres}}|$.

\subsection{The approach followed in our previous studies}\label{trad}
 It has been repeatedly confirmed\cite{NAT01,NAT02A,VAR05C,NAT06A,NAT06B,JAP08} that,
 when an SES activity is observed, one can specify an area A by the selectivity map and (at least) one magnitude threshold $M_{thres}$ which satisfy the conditions (\ref{ex3}) and (\ref{eq2}), i.e.,$\kappa_1({\cal A}_{{\rm M}_{thres}})\approx 0.07$  and $S({\cal A}_{{\rm M}_{thres}}), S_-({\cal A}_{{\rm M}_{thres}})< S_u $, a few days to a few days before the main shock.  Thus, such a study enables, in principle, prediction of the main shock to be made within a few days to a few hours before its occurrence.

   The actual procedure was carried out as follows: For seismicity analysis, we set the natural time zero at the initiation of the SES, and then formed time series of seismic events in natural time each time when (small) EQs in A occurred, namely the number $j$ increased. 
   The normalized power spectrum in natural time  for each of the time series was computed from the pairs $(\chi_k,Q_k)=(k/j, M_0(e_k)), k=1,\ldots j$, and compared with that of Eq.(\ref{fasma}) for $\phi \in [0,0.5]$. It was found  that $\Pi (\phi )$
approaches that of Eq.(\ref{fasma}) a few hours to a few days before the main shock.  We also calculated the evolution of the quantities $\kappa_1$, $S$ and $S_{-}$ to ascertain Eq. (\ref{eq2}) was also satisfied. The conditions for a true coincidence of observed time series with that of critical state were adopted as follows \cite{NAT01,newbook,VAR05C,NAT06A,NAT06B,JAP08}: 
First, the
`average' distance $\langle D \rangle$ between the  $\Pi(\phi )$ of the evolving seismicity  and that of Eq.(\ref{fasma}) should be smaller than $10^{-2}$. This was a practical criterion. 
Second, the final approach of $\kappa_1$  of the evolving $\Pi(\phi )$ to that of Eq.(\ref{fasma}), i.e., 0.07, must be by descending from above. This condition 
was found empirically. Third, both
values $S$ and $S_{-}$ should be smaller than $S_u$ at the
coincidence. Finally and fourth, since the process concerned is self-similar
({\em critical} dynamics), the time of the occurrence of the
(true) coincidence should {\em not} vary, in principle, upon
changing (within reasonable limits) the  the size of area and the magnitude threshold $M_{thres}$.

\subsection{A new approach suggested here upon using  ${\cal P}_{{\rm M}_{thres}}$  or the ${\cal E[ A}_{{\rm M}_{thres}}]$ ensemble}

 The basic idea behind the new approach suggested in this paper is the following:
  When area A reaches criticality, one expects in general that all 
  its subareas have also reached criticality {\em simultaneously}.
   Each of these subareas corresponds to a certain value of $\kappa_1$ since
  the events that sequentially occur in that subarea (after the initiation of the SES activity) constitute a time-ordered set (which is a subset of ${\cal A}_{{\rm M}_{thres}}$), the natural time analysis of which results in a certain $\kappa_1$-value. Thus, we expect that the distribution of these $\kappa_1$-values obtained from the analysis of all subareas should be centered around 0.070, according to Eq.(\ref{ex3}).
  
In principle, in order to investigate whether criticality has been approached immediately after the occurrence of a new event $i$ within the predicted area A, we should construct all possible subsets that {\em necessarily} include the event $i$. Each of these subsets may be considered as a proper subset (${\cal P}_{{\rm M}_{thres}}$) if and only if it includes all EQs that took place inside its corresponding rectangular subarea, i.e., $R({\cal P}_{{\rm M}_{thres}})$. (This is, of course an approximation -followed through out the present paper- because other geometries, e.g., circular, could be also considered.) In what follows, we will solely focus on such proper subsets of ${\cal A}_{{\rm M}_{thres}}$.

\begin{figure}
\includegraphics{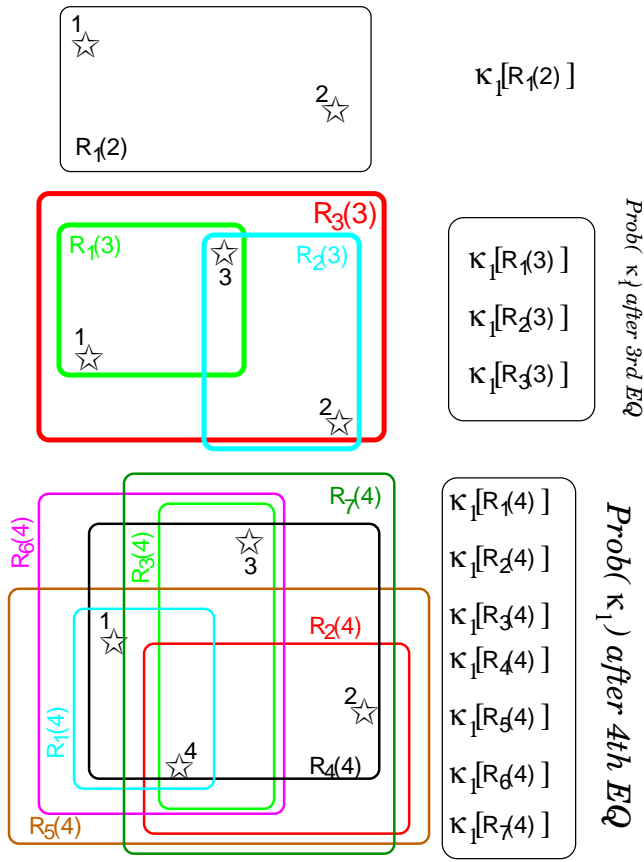}
\caption{The proper subsets and the corresponding rectangular subareas immediately after the occurrence of the second earthquake ``2'' (upper panel), the third earthquake ``3'' (middle panel) and the fourth earthquake ``4'' (bottom panel). The location of each earthquake is shown by an open star.}\label{sx1}
\end{figure}

Let us now consider an example in which four earthquakes shown in Fig.\ref{sx1} have occurred in the area A, in a sequence indicated by the numbers 1,2,3 and 4. Figure \ref{sx1} depicts the proper subsets of ${\cal A}_{{\rm M}_{thres}}$ just after the occurrence of each earthquake. In these proper subsets (which form the ${\cal E[ A}_{{\rm M}_{thres}}]$  ensemble at each time instant), one has to compute the corresponding $\kappa_1$ values and then construct their distribution Prob($\kappa_1$). The latter distribution depicts the state of the ensemble ${\cal E[ A}_{{\rm M}_{thres}}]$. For example, just after the occurrence of the second event a single proper subset can be defined, thus only $\kappa_1[R_1(2)]$ is available. Just after the occurrence of the third event, three proper subsets of ${\cal A}_{{\rm M}_{thres}}$ can be defined as shown in Fig. \ref{sx1}. Recall that the necessary condition for having a proper subset at a given time instant is that it should include the last event (the third earthquake in this case) and for this reason the result corresponding to $R_1(2)$ is not considered for the construction of the distribution Prob($\kappa_1$) [at this time instant]. By the same token, after the occurrence of the fourth event, seven proper subsets result, that can be visualized in Fig. \ref{sx1}. Thus, we can now calculate $\kappa_1$ for each of these 7 subsets and –after assuming equipartition of probability among these subsets- we construct the Prob($\kappa_1$) versus $\kappa_1$ graph and then examine whether it maximizes at $\kappa_1\approx$ 0.070 (i.e., obeys Eq.(\ref{ex3})).

In other words, in the new approach, the $\kappa_1$-values of all these subareas that enclose earthquakes that occurred after the initiation of the SES until a given time instant are included together with the corresponding results of the largest area A, all of them treated on {\em equal} footing. Then by performing an averaging procedure over {\em all} those proper subsets (which correspond, at a given time instant, to a large number of subareas of A), we expect that the average will also satisfy Eq.(\ref{ex3}).

We shall demonstrate in the next section by using precise experimental examples, which refer to the most recent SES activities recorded in Greece, that the following may also determine the time-window of the impending main shock: 
The average value $\langle \kappa_1({\cal P}_{{\rm M}_{thres}} \rangle_{{\cal E[ A}_{{\rm M}_{thres}}]}$ obtained 
when using the ensemble ${\cal E[ A}_{{\rm M}_{thres}}]$  satisfies the condition
(\ref{ex3}) and then the mainshock occurs within a few days at the most.

\begin{figure}
\includegraphics{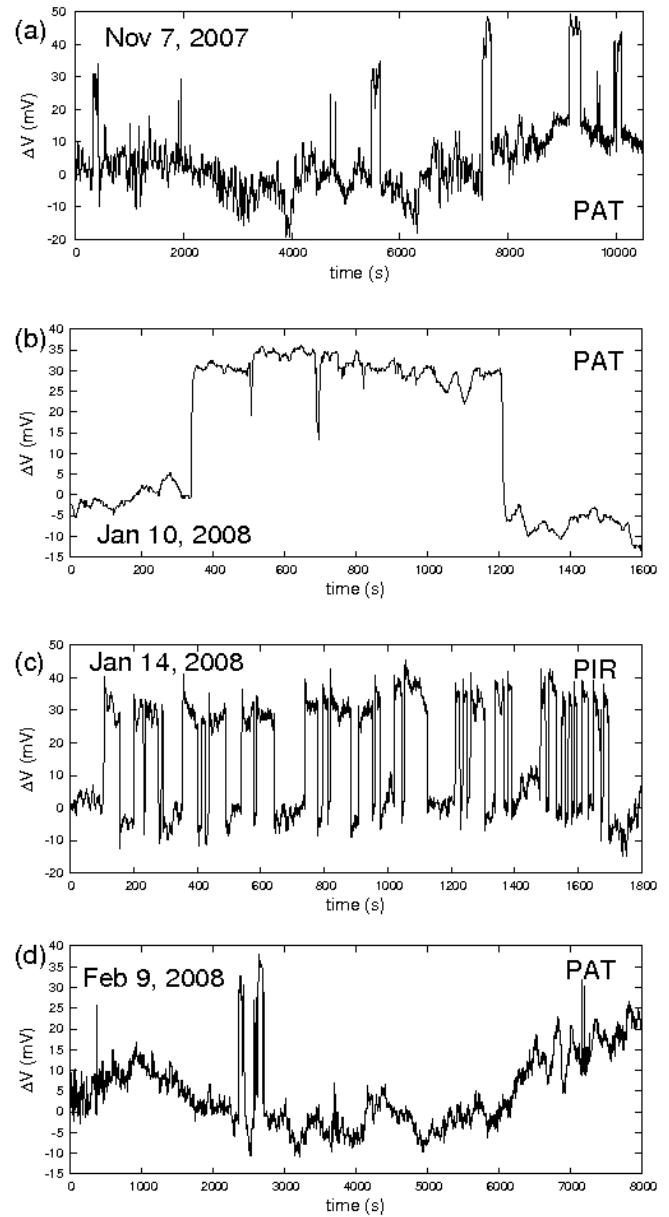}
\caption{The most recent SES activities recorded in Greece.} \label{SESact}
\end{figure}

\section{The application of the proposed procedure to the most recent examples}
\label{tria}

In Fig.\ref{SESact}, we depict four SES activities that have been recently recorded at the stations 
PAT (in central Greece) and PIR (western Greece): (a) on November 7, 2007, at PAT (b) 
on January 10, 2008, at PAT (c) on January 14, 2008, at PIR and (d) on February 9, 2008, at PAT. These have been classified as SES activities by applying the usual criteria (i.e., the conditions \ref{ex3} and \ref{eq2}) explained in detail in 
Ref.\cite{vers1,vers3,SAR08A}. In addition, two long duration SES activities were recorded at PIR from January 21 to January 26, 2008 and from February 29 to March 2, 2008 that will be described separately in Subsection E. We now apply the present procedure to all these cases:

\subsection{The case of the SES activity on Nov. 7, 2007}
The investigation of the seismicity subsequent to that SES activity was made in the area A:$N_{37.6}^{38.6}E_{20.0}^{23.3}$ which was already indicated in Ref.\cite{vers1}. 
At 05:14 UT on 6th January, 2008, a major earthquake (M6.6) occurred with epicenter located at 37.1$^o$N22.8$^o$E, i.e., only around 50km to the south of the area A studied. What happened before that EQ can be seen in Fig.\ref{Leo}. An inspection of this figure reveals that the Prob($\kappa_1$) maximizes at around $\kappa_1=0.073$ and $\langle \kappa_1({\cal P}_{{\rm M}_{thres}} \rangle_{{\cal E[ A}_{{\rm M}_{thres}}]}=0.070$ with standard deviation 0.008 upon the occurrence of a small event at 04:32 on January 4, 2008, i.e., almost two days before the main shock. 

\subsection{The case of the SES activity on Jan. 10, 2008}
The investigation of the seismicity was also made\cite{vers3} in the area A:$N_{37.6}^{38.6}E_{20.0}^{23.3}$.  The results are shown in Fig.\ref{Pat}, where we see that  Prob($\kappa_1$) exhibits bimodal feature with a secondary peak at  $\kappa_1\approx0.070$  upon the occurrence of the small events at 12:21 UT and 13:26 UT on February 3, 2008 (these two cases are shown with arrows). Actually,  at 20:25 UT and 22:15 UT on February 4, 2008 two EQs with magnitudes 5.4 and 5.5 occurred with epicenters around 38.1$^o$N21.9$^o$E lying at a small distance ($\approx$10km) from the measuring station PAT and inside the area studied.

\subsection{The case of the SES activity at PIR on Jan. 14, 2008}
Since this SES activity was recorded at PIR, the study of  the seismicity was made in the area $N_{36.0}^{38.6}E_{20.0}^{22.5}$ indicated in Ref.\cite{vers3} well in advance. The results of the computation are depicted in Fig.\ref{p14}, which reveals that the Prob($\kappa_1$)  also exhibits bimodal feature, one mode of which 
maximizes at $\kappa_1\approx0.07$ upon the occurrence of a small event at 04:07 UT on February 12, 2008. Almost two days later, i.e., at 10:09 UT on February 14, 2008, a major EQ of magnitude 6.7 occurred at 36.5$^o$N21.8$^o$E. This EQ -according to USGS calatogue (which reported $M_w$6.9)- is the strongest EQ that occurred in Greece during the last twenty years. 
In addition, a few hours later, i.e., at 12:08 UT, a M6.6 earthquake occurred at 36.2$^o$N 21.8$^o$E.

\subsection{The case of the SES activity at PAT on Feb. 9, 2008}

The  results until early in the morning on March 19, 2008, of the study of the seismicity in the area\cite{SAR08A} $N_{37.5}^{38.6}E_{20.0}^{23.3}$ can be visualized in Fig.
\ref{oldfig5} which shows that Prob($\kappa_1$) has not exhibited a maximum at $\kappa_1\approx0.07$ yet.  This investigation is still in progress.

\subsection{The case of the SES activity at PIR during the period February 29, 2008 to March 2, 2008 (Fig.\ref{fig5}(c))}

This was a long duration SES activity (see Fig. \ref{fig5}(c)) with polarity opposite to that of another long duration SES activity which was observed also at PIR from January 21, 2008 to January 26, 2008 (Fig.\ref{fig5}(b)). Further, for the sake of comparison, in Fig.\ref{fig5}(a), we present one more long duration SES activity on September 17, 2005 at PIR that was followed by the Mw6.7 EQ at 36.3$^o$N23.2$^o$E on January 8, 2006. The occurrence of the latter two major EQs (Jan.8, 2006 and Feb.14, 2008), leads to the updated selectivity map of PIR as shown by the shaded area in Fig.\ref{fig6} which lies along the Hellenic arc (marked with a thick solid line). In other words, the impending main shock for the SES activity of February 29, 2008 to March 2, 2008 is expected to occur in this shaded area.

A further study of the imminent seismicity is now in progress in order to clarify which region of the shaded area in Fig.\ref{fig6}, will finally exhibit the expected feature ,i.e., the maximization of Prob($\kappa_1$) at $\kappa_1 \approx$0.07 (see Appendix).

\section{Conclusion}
\label{tesera}
Upon the recording of an SES activity, one can estimate an area A within which the impending mainshock is expected to occur. Following the 
subsequent seismicity, the probability density function of $\kappa_1$ is obtained, which maximizes at $\kappa_1 \approx 0.070$ a few days at the most before the occurrence of the mainshock.

\appendix*
\section{Earthquakes that followed the long duration SES activity at PIR from  February 29, 2008 to March 2, 2008}

At 19:16 UT on March 25, 2008, the ongoing investigation of the seismicity (for $M_{thres}=3.2$) in the area $N_{37.0}^{38.6}E_{20.0}^{22.0}$ (see Ref.\cite{SAR08A}), after the SES activity of February 29 to March 2, 2008 at PIR, showed a maximization of Prob($\kappa_1$) at $\kappa_1\approx0.070$ as shown in Fig.\ref{figa1}. 
This was forwarded to various interested Institutes in Europe, Japan and USA at 21:47 UT on March 26, 2008. Actually, almost one day later, i.e., at 00:16 UT on March 28, 2008, a Ms(ATH)=5.7 EQ (PDE of USGS reported Mw=5.6), occurred at 35.0$^o$N 25.3$^o$E lying approximately 150 km to the east of the southern part of the PIR selectivity map shown by the shaded area in Fig.\ref{fig6}.

In view of the amplitude of the SES activity in Fig. \ref{fig5}(c) discussed in the main text (which is comparable to that of the SES activities depicted in Figs.\ref{fig5}(a) and \ref{fig5}(b)), the study of the seismicity still continued to investigate if a more pronounced peak of Prob($\kappa_1$) at $\kappa_1\approx$0.070 would eventually occur in the near future. This actually happened at 12:26 UT on May 8, 2008 and almost two days later, 
i.e., at 20:53 UT on May 10, 2008 a Ms(ATH)=5.6 EQ occurred with an epicenter at 
36.4$^o$N22.3$^o$E (see also below). 

\subsection{Note Added on May 19, 2008.} In continuation of the study mentioned in the Appendix, the ongoing investigation of the seismicity in the shaded area of Fig.\ref{fig6} reveals the following (cf. A calculation with $M_{thres}=3.2$ similar to the one in the previous cases, cannot be practically carried out in this case due to the large number of the events, i.e., more than half a thousand, involved in the calculation; recall that $M_{thres}$ refer to 
the ML values reported by the Athens Observatory):
 
{\bf (i)}For $M_{thres}=3.8$, a maximization of Prob($\kappa_1$) at $\kappa_1 \approx$0.070 was observed at 00:59 UT on May 16, 2008 upon the occurrence of a ML=3.8 event at 36.1$^o$N 21.8$^o$E.

{\bf (ii)} For $M_{thres}=3.7$, Prob($\kappa_1$) was clearly maximized at $\kappa_1 \approx$0.070 when a ML=3.8 event occurred at 36.6$^o$N 20.9$^o$E at 20:27 UT on May 18, 2008.

{\bf (iii)} For $M_{thres}=3.6$, Prob($\kappa_1$) exhibited a maximum at $\kappa_1 \approx$0.070 at 10:28 UT on May 18, 2008 upon the occurrence of a ML=3.8 event at 35.9$^o$N 23.3$^o$E.

The aforementioned calculations show that, at a first approximation and within reasonable time limits, magnitude threshold invariance seems to hold. The extent to which, this behaviour will conform to the main conclusion of the present paper, it remains to be seen.

\subsection{Note Added on May 29, 2008.} At 15:11 UT on May 25, 2008 a Ms(ATH)=4.6 EQ occurred 
at 38.2$^o$N 22.7$^o$E (PDE of USGS reported mb=4.7). It is not yet clear whether this EQ 
is associated with the shaded area in Fig.\ref{fig6} (in particular with its lobe that almost reaches PAT) or with the rectangular area   $N_{37.5}^{38.6}E_{20.0}^{23.3}$. 

In addition, at 23:26 UT on May 27, 2008 a Ms(ATH)=5.1 EQ occurred with an epicenter 
around 35.5$^o$N 22.4$^o$E, as expected by the Note added on May 19, 2008. Upon the occurrence of this event,  Prob($\kappa_1$) exhibits 
a pronounced maximum at $\kappa_1\approx 0.07$ marked by an arrow in Fig.\ref{figa2} drawn for $M_{thres}=3.9$. (An additional arrow marks an earlier maximum on
May  8, 2008 that preceded the aforementioned  Ms(ATH)=5.6 EQ on May 10, 2008). Quite interestingly, this exhibits magnitude threshold invariance (a behavior that should be obeyed 
at the {\em critical} point) since a similar maximum at $\kappa_1=0.07$  appears {\em simultaneously } for $M_{thres}=4.0$ and  $M_{thres}=4.1$ as can be verified by an inspection 
of Figs.(\ref{figa3}) and (\ref{figa4}), respectively.






\begin{thebibliography}{48}
\expandafter\ifx\csname natexlab\endcsname\relax\def\natexlab#1{#1}\fi
\expandafter\ifx\csname bibnamefont\endcsname\relax
  \def\bibnamefont#1{#1}\fi
\expandafter\ifx\csname bibfnamefont\endcsname\relax
  \def\bibfnamefont#1{#1}\fi
\expandafter\ifx\csname citenamefont\endcsname\relax
  \def\citenamefont#1{#1}\fi
\expandafter\ifx\csname url\endcsname\relax
  \def\url#1{\texttt{#1}}\fi
\expandafter\ifx\csname urlprefix\endcsname\relax\def\urlprefix{URL }\fi
\providecommand{\bibinfo}[2]{#2}
\providecommand{\eprint}[2][]{\url{#2}}

\bibitem[{\citenamefont{Varotsos and Alexopoulos}(1984{\natexlab{a}})}]{proto}
\bibinfo{author}{\bibfnamefont{P.}~\bibnamefont{Varotsos}} \bibnamefont{and}
  \bibinfo{author}{\bibfnamefont{K.}~\bibnamefont{Alexopoulos}},
  \bibinfo{journal}{Tectonophysics} \textbf{\bibinfo{volume}{110}},
  \bibinfo{pages}{73} (\bibinfo{year}{1984}{\natexlab{a}}).

\bibitem[{\citenamefont{Varotsos and Alexopoulos}(1984{\natexlab{b}})}]{proto2}
\bibinfo{author}{\bibfnamefont{P.}~\bibnamefont{Varotsos}} \bibnamefont{and}
  \bibinfo{author}{\bibfnamefont{K.}~\bibnamefont{Alexopoulos}},
  \bibinfo{journal}{Tectonophysics} \textbf{\bibinfo{volume}{110}},
  \bibinfo{pages}{99} (\bibinfo{year}{1984}{\natexlab{b}}).

\bibitem[{\citenamefont{Varotsos et~al.}(1988)\citenamefont{Varotsos,
  Alexopoulos, Nomicos, and Lazaridou}}]{var88x}
\bibinfo{author}{\bibfnamefont{P.}~\bibnamefont{Varotsos}},
  \bibinfo{author}{\bibfnamefont{K.}~\bibnamefont{Alexopoulos}},
  \bibinfo{author}{\bibfnamefont{K.}~\bibnamefont{Nomicos}}, \bibnamefont{and}
  \bibinfo{author}{\bibfnamefont{M.}~\bibnamefont{Lazaridou}},
  \bibinfo{journal}{Tectonophysics} \textbf{\bibinfo{volume}{152}},
  \bibinfo{pages}{193} (\bibinfo{year}{1988}).

\bibitem[{\citenamefont{Varotsos and Lazaridou}(1991)}]{VAR91}
\bibinfo{author}{\bibfnamefont{P.}~\bibnamefont{Varotsos}} \bibnamefont{and}
  \bibinfo{author}{\bibfnamefont{M.}~\bibnamefont{Lazaridou}},
  \bibinfo{journal}{Tectonophysics} \textbf{\bibinfo{volume}{188}},
  \bibinfo{pages}{321} (\bibinfo{year}{1991}).

\bibitem[{\citenamefont{Varotsos et~al.}(1993)\citenamefont{Varotsos,
  Alexopoulos, and Lazaridou}}]{VAR93}
\bibinfo{author}{\bibfnamefont{P.}~\bibnamefont{Varotsos}},
  \bibinfo{author}{\bibfnamefont{K.}~\bibnamefont{Alexopoulos}},
  \bibnamefont{and}
  \bibinfo{author}{\bibfnamefont{M.}~\bibnamefont{Lazaridou}},
  \bibinfo{journal}{Tectonophysics} \textbf{\bibinfo{volume}{224}},
  \bibinfo{pages}{1} (\bibinfo{year}{1993}).

\bibitem[{\citenamefont{Varotsos et~al.}(1999)\citenamefont{Varotsos, Sarlis,
  and Lazaridou}}]{var99}
\bibinfo{author}{\bibfnamefont{P.~V.} \bibnamefont{Varotsos}},
  \bibinfo{author}{\bibfnamefont{N.~V.} \bibnamefont{Sarlis}},
  \bibnamefont{and} \bibinfo{author}{\bibfnamefont{M.~S.}
  \bibnamefont{Lazaridou}}, \bibinfo{journal}{Phys. Rev. B}
  \textbf{\bibinfo{volume}{59}}, \bibinfo{pages}{24} (\bibinfo{year}{1999}).

\bibitem[{\citenamefont{Sarlis et~al.}(1999)\citenamefont{Sarlis, Lazaridou,
  Kapiris, and Varotsos}}]{grl}
\bibinfo{author}{\bibfnamefont{N.}~\bibnamefont{Sarlis}},
  \bibinfo{author}{\bibfnamefont{M.}~\bibnamefont{Lazaridou}},
  \bibinfo{author}{\bibfnamefont{P.}~\bibnamefont{Kapiris}}, \bibnamefont{and}
  \bibinfo{author}{\bibfnamefont{P.}~\bibnamefont{Varotsos}},
  \bibinfo{journal}{Geophys. Res. Lett.} \textbf{\bibinfo{volume}{26}},
  \bibinfo{pages}{3245} (\bibinfo{year}{1999}).

\bibitem[{\citenamefont{Uyeda et~al.}(2000)\citenamefont{Uyeda, Nagao, Orihara,
  Yamaguchi, and Takahashi}}]{uye}
\bibinfo{author}{\bibfnamefont{S.}~\bibnamefont{Uyeda}},
  \bibinfo{author}{\bibfnamefont{T.}~\bibnamefont{Nagao}},
  \bibinfo{author}{\bibfnamefont{Y.}~\bibnamefont{Orihara}},
  \bibinfo{author}{\bibfnamefont{T.}~\bibnamefont{Yamaguchi}},
  \bibnamefont{and}
  \bibinfo{author}{\bibfnamefont{I.}~\bibnamefont{Takahashi}},
  \bibinfo{journal}{Proc. Natl. Acad. Sci. USA} \textbf{\bibinfo{volume}{97}},
  \bibinfo{pages}{4561} (\bibinfo{year}{2000}).

\bibitem[{\citenamefont{Uyeda et~al.}(2002)\citenamefont{Uyeda, Hayakawa,
  Nagao, Molchanov, Hattori, Orihara, Gotoh, Akinaga, and Tanaka}}]{uye2}
\bibinfo{author}{\bibfnamefont{S.}~\bibnamefont{Uyeda}},
  \bibinfo{author}{\bibfnamefont{M.}~\bibnamefont{Hayakawa}},
  \bibinfo{author}{\bibfnamefont{T.}~\bibnamefont{Nagao}},
  \bibinfo{author}{\bibfnamefont{O.}~\bibnamefont{Molchanov}},
  \bibinfo{author}{\bibfnamefont{K.}~\bibnamefont{Hattori}},
  \bibinfo{author}{\bibfnamefont{Y.}~\bibnamefont{Orihara}},
  \bibinfo{author}{\bibfnamefont{K.}~\bibnamefont{Gotoh}},
  \bibinfo{author}{\bibfnamefont{Y.}~\bibnamefont{Akinaga}}, \bibnamefont{and}
  \bibinfo{author}{\bibfnamefont{H.}~\bibnamefont{Tanaka}},
  \bibinfo{journal}{Proc. Natl. Acad. Sci. USA} \textbf{\bibinfo{volume}{99}},
  \bibinfo{pages}{7352} (\bibinfo{year}{2002}).

\bibitem[{\citenamefont{Flores-M\'{a}rquez
  et~al.}(2007)\citenamefont{Flores-M\'{a}rquez, M\'{a}rquez-Cruz,
  Ramirez-Rojas, G\'{a}lvez-Coyt, and Angulo-Brown}}]{MEX07}
\bibinfo{author}{\bibfnamefont{L.}~\bibnamefont{Flores-M\'{a}rquez}},
  \bibinfo{author}{\bibfnamefont{J.}~\bibnamefont{M\'{a}rquez-Cruz}},
  \bibinfo{author}{\bibfnamefont{A.}~\bibnamefont{Ramirez-Rojas}},
  \bibinfo{author}{\bibfnamefont{G.}~\bibnamefont{G\'{a}lvez-Coyt}},
  \bibnamefont{and}
  \bibinfo{author}{\bibfnamefont{F.}~\bibnamefont{Angulo-Brown}},
  \bibinfo{journal}{Nat. Hazards Earth Syst. Sci.}
  \textbf{\bibinfo{volume}{7}}, \bibinfo{pages}{549} (\bibinfo{year}{2007}).

\bibitem[{\citenamefont{Varotsos and Alexopoulos}(1986)}]{varbook}
\bibinfo{author}{\bibfnamefont{P.}~\bibnamefont{Varotsos}} \bibnamefont{and}
  \bibinfo{author}{\bibfnamefont{K.}~\bibnamefont{Alexopoulos}},
  \emph{\bibinfo{title}{Thermodynamics of Point Defects and their Relation with
  Bulk Properties}} (\bibinfo{publisher}{North Holland},
  \bibinfo{address}{Amsterdam}, \bibinfo{year}{1986}).

\bibitem[{\citenamefont{Varotsos}(2005)}]{newbook}
\bibinfo{author}{\bibfnamefont{P.}~\bibnamefont{Varotsos}},
  \emph{\bibinfo{title}{The Physics of Seismic Electric Signals}}
  (\bibinfo{publisher}{TERRAPUB}, \bibinfo{address}{Tokyo},
  \bibinfo{year}{2005}).

\bibitem[{\citenamefont{Varotsos}(1977)}]{VAR77}
\bibinfo{author}{\bibfnamefont{P.}~\bibnamefont{Varotsos}},
  \bibinfo{journal}{J. Physique(Paris) Lettr.} \textbf{\bibinfo{volume}{38}},
  \bibinfo{pages}{L455} (\bibinfo{year}{1977}).

\bibitem[{\citenamefont{Varotsos and Alexopoulos}(1978)}]{VAR78}
\bibinfo{author}{\bibfnamefont{P.}~\bibnamefont{Varotsos}} \bibnamefont{and}
  \bibinfo{author}{\bibfnamefont{K.}~\bibnamefont{Alexopoulos}},
  \bibinfo{journal}{Phys. Stat. Solidi A} \textbf{\bibinfo{volume}{47}},
  \bibinfo{pages}{K133} (\bibinfo{year}{1978}).

\bibitem[{\citenamefont{Lazaridou et~al.}(1985)\citenamefont{Lazaridou,
  Varotsos, Alexopoulos, and Varotsos}}]{VAR85}
\bibinfo{author}{\bibfnamefont{M.}~\bibnamefont{Lazaridou}},
  \bibinfo{author}{\bibfnamefont{C.}~\bibnamefont{Varotsos}},
  \bibinfo{author}{\bibfnamefont{K.}~\bibnamefont{Alexopoulos}},
  \bibnamefont{and} \bibinfo{author}{\bibfnamefont{P.}~\bibnamefont{Varotsos}},
  \bibinfo{journal}{J. Phys. C: Solid State} \textbf{\bibinfo{volume}{18}},
  \bibinfo{pages}{3891} (\bibinfo{year}{1985}).

\bibitem[{\citenamefont{Varotsos
  et~al.}(2001{\natexlab{a}})\citenamefont{Varotsos, Sarlis, and
  Skordas}}]{PJA1}
\bibinfo{author}{\bibfnamefont{P.}~\bibnamefont{Varotsos}},
  \bibinfo{author}{\bibfnamefont{N.}~\bibnamefont{Sarlis}}, \bibnamefont{and}
  \bibinfo{author}{\bibfnamefont{E.}~\bibnamefont{Skordas}},
  \bibinfo{journal}{Proc. Jpn. Acad., Ser. B: Phys. Biol. Sci.}
  \textbf{\bibinfo{volume}{77}}, \bibinfo{pages}{87}
  (\bibinfo{year}{2001}{\natexlab{a}}).

\bibitem[{\citenamefont{Varotsos
  et~al.}(2001{\natexlab{b}})\citenamefont{Varotsos, Sarlis, and
  Skordas}}]{PJA2}
\bibinfo{author}{\bibfnamefont{P.}~\bibnamefont{Varotsos}},
  \bibinfo{author}{\bibfnamefont{N.}~\bibnamefont{Sarlis}}, \bibnamefont{and}
  \bibinfo{author}{\bibfnamefont{E.}~\bibnamefont{Skordas}},
  \bibinfo{journal}{Proc. Jpn. Acad., Ser. B: Phys. Biol. Sci.}
  \textbf{\bibinfo{volume}{77}}, \bibinfo{pages}{93}
  (\bibinfo{year}{2001}{\natexlab{b}}).

\bibitem[{\citenamefont{Varotsos
  et~al.}(2003{\natexlab{a}})\citenamefont{Varotsos, Sarlis, and
  Skordas}}]{PRL03}
\bibinfo{author}{\bibfnamefont{P.~A.} \bibnamefont{Varotsos}},
  \bibinfo{author}{\bibfnamefont{N.~V.} \bibnamefont{Sarlis}},
  \bibnamefont{and} \bibinfo{author}{\bibfnamefont{E.~S.}
  \bibnamefont{Skordas}}, \bibinfo{journal}{Phys. Rev. Lett.}
  \textbf{\bibinfo{volume}{91}}, \bibinfo{pages}{148501}
  (\bibinfo{year}{2003}{\natexlab{a}}).

\bibitem[{\citenamefont{Varotsos
  et~al.}(2002{\natexlab{a}})\citenamefont{Varotsos, Sarlis, and
  Skordas}}]{NAT02}
\bibinfo{author}{\bibfnamefont{P.~A.} \bibnamefont{Varotsos}},
  \bibinfo{author}{\bibfnamefont{N.~V.} \bibnamefont{Sarlis}},
  \bibnamefont{and} \bibinfo{author}{\bibfnamefont{E.~S.}
  \bibnamefont{Skordas}}, \bibinfo{journal}{Phys. Rev. E}
  \textbf{\bibinfo{volume}{66}}, \bibinfo{pages}{011902}
  (\bibinfo{year}{2002}{\natexlab{a}}).

\bibitem[{\citenamefont{Weron et~al.}(2005)\citenamefont{Weron, Burnecki,
  Mercik, and Weron}}]{WER05}
\bibinfo{author}{\bibfnamefont{A.}~\bibnamefont{Weron}},
  \bibinfo{author}{\bibfnamefont{K.}~\bibnamefont{Burnecki}},
  \bibinfo{author}{\bibfnamefont{S.}~\bibnamefont{Mercik}}, \bibnamefont{and}
  \bibinfo{author}{\bibfnamefont{K.}~\bibnamefont{Weron}},
  \bibinfo{journal}{Phys. Rev. E} \textbf{\bibinfo{volume}{71}},
  \bibinfo{pages}{016113} (\bibinfo{year}{2005}).

\bibitem[{\citenamefont{Varotsos
  et~al.}(2001{\natexlab{c}})\citenamefont{Varotsos, Sarlis, and
  Skordas}}]{NAT01}
\bibinfo{author}{\bibfnamefont{P.~A.} \bibnamefont{Varotsos}},
  \bibinfo{author}{\bibfnamefont{N.~V.} \bibnamefont{Sarlis}},
  \bibnamefont{and} \bibinfo{author}{\bibfnamefont{E.~S.}
  \bibnamefont{Skordas}}, \bibinfo{journal}{Practica of Athens Academy}
  \textbf{\bibinfo{volume}{76}}, \bibinfo{pages}{294}
  (\bibinfo{year}{2001}{\natexlab{c}}).

\bibitem[{\citenamefont{Varotsos
  et~al.}(2002{\natexlab{b}})\citenamefont{Varotsos, Sarlis, and
  Skordas}}]{NAT02A}
\bibinfo{author}{\bibfnamefont{P.~A.} \bibnamefont{Varotsos}},
  \bibinfo{author}{\bibfnamefont{N.~V.} \bibnamefont{Sarlis}},
  \bibnamefont{and} \bibinfo{author}{\bibfnamefont{E.~S.}
  \bibnamefont{Skordas}}, \bibinfo{journal}{Acta Geophys. Pol.}
  \textbf{\bibinfo{volume}{50}}, \bibinfo{pages}{337}
  (\bibinfo{year}{2002}{\natexlab{b}}).

\bibitem[{\citenamefont{Varotsos
  et~al.}(2005{\natexlab{a}})\citenamefont{Varotsos, Sarlis, Tanaka, and
  Skordas}}]{VAR05C}
\bibinfo{author}{\bibfnamefont{P.~A.} \bibnamefont{Varotsos}},
  \bibinfo{author}{\bibfnamefont{N.~V.} \bibnamefont{Sarlis}},
  \bibinfo{author}{\bibfnamefont{H.~K.} \bibnamefont{Tanaka}},
  \bibnamefont{and} \bibinfo{author}{\bibfnamefont{E.~S.}
  \bibnamefont{Skordas}}, \bibinfo{journal}{Phys. Rev. E}
  \textbf{\bibinfo{volume}{72}}, \bibinfo{pages}{041103}
  (\bibinfo{year}{2005}{\natexlab{a}}).

\bibitem[{\citenamefont{Varotsos
  et~al.}(2006{\natexlab{a}})\citenamefont{Varotsos, Sarlis, Skordas, Tanaka,
  and Lazaridou}}]{NAT06A}
\bibinfo{author}{\bibfnamefont{P.~A.} \bibnamefont{Varotsos}},
  \bibinfo{author}{\bibfnamefont{N.~V.} \bibnamefont{Sarlis}},
  \bibinfo{author}{\bibfnamefont{E.~S.} \bibnamefont{Skordas}},
  \bibinfo{author}{\bibfnamefont{H.~K.} \bibnamefont{Tanaka}},
  \bibnamefont{and} \bibinfo{author}{\bibfnamefont{M.~S.}
  \bibnamefont{Lazaridou}}, \bibinfo{journal}{Phys. Rev. E}
  \textbf{\bibinfo{volume}{73}}, \bibinfo{pages}{031114}
  (\bibinfo{year}{2006}{\natexlab{a}}).

\bibitem[{\citenamefont{Varotsos
  et~al.}(2006{\natexlab{b}})\citenamefont{Varotsos, Sarlis, Skordas, Tanaka,
  and Lazaridou}}]{NAT06B}
\bibinfo{author}{\bibfnamefont{P.~A.} \bibnamefont{Varotsos}},
  \bibinfo{author}{\bibfnamefont{N.~V.} \bibnamefont{Sarlis}},
  \bibinfo{author}{\bibfnamefont{E.~S.} \bibnamefont{Skordas}},
  \bibinfo{author}{\bibfnamefont{H.~K.} \bibnamefont{Tanaka}},
  \bibnamefont{and} \bibinfo{author}{\bibfnamefont{M.~S.}
  \bibnamefont{Lazaridou}}, \bibinfo{journal}{Phys. Rev. E}
  \textbf{\bibinfo{volume}{74}}, \bibinfo{pages}{021123}
  (\bibinfo{year}{2006}{\natexlab{b}}).

\bibitem[{\citenamefont{Varotsos
  et~al.}(2008{\natexlab{a}})\citenamefont{Varotsos, Sarlis, Skordas, and
  Lazaridou}}]{JAP08}
\bibinfo{author}{\bibfnamefont{P.~A.} \bibnamefont{Varotsos}},
  \bibinfo{author}{\bibfnamefont{N.~V.} \bibnamefont{Sarlis}},
  \bibinfo{author}{\bibfnamefont{E.~S.} \bibnamefont{Skordas}},
  \bibnamefont{and} \bibinfo{author}{\bibfnamefont{M.~S.}
  \bibnamefont{Lazaridou}}, \bibinfo{journal}{J. Appl. Phys.}
  \textbf{\bibinfo{volume}{103}}, \bibinfo{pages}{014906}
  (\bibinfo{year}{2008}{\natexlab{a}}).

\bibitem[{\citenamefont{Varotsos
  et~al.}(2003{\natexlab{b}})\citenamefont{Varotsos, Sarlis, and
  Skordas}}]{NAT03}
\bibinfo{author}{\bibfnamefont{P.~A.} \bibnamefont{Varotsos}},
  \bibinfo{author}{\bibfnamefont{N.~V.} \bibnamefont{Sarlis}},
  \bibnamefont{and} \bibinfo{author}{\bibfnamefont{E.~S.}
  \bibnamefont{Skordas}}, \bibinfo{journal}{Phys. Rev. E}
  \textbf{\bibinfo{volume}{67}}, \bibinfo{pages}{021109}
  (\bibinfo{year}{2003}{\natexlab{b}}).

\bibitem[{\citenamefont{Varotsos
  et~al.}(2003{\natexlab{c}})\citenamefont{Varotsos, Sarlis, and
  Skordas}}]{NAT03B}
\bibinfo{author}{\bibfnamefont{P.~A.} \bibnamefont{Varotsos}},
  \bibinfo{author}{\bibfnamefont{N.~V.} \bibnamefont{Sarlis}},
  \bibnamefont{and} \bibinfo{author}{\bibfnamefont{E.~S.}
  \bibnamefont{Skordas}}, \bibinfo{journal}{Phys. Rev. E}
  \textbf{\bibinfo{volume}{68}}, \bibinfo{pages}{031106}
  (\bibinfo{year}{2003}{\natexlab{c}}).

\bibitem[{\citenamefont{Varotsos et~al.}(2004)\citenamefont{Varotsos, Sarlis,
  Skordas, and Lazaridou}}]{NAT04}
\bibinfo{author}{\bibfnamefont{P.~A.} \bibnamefont{Varotsos}},
  \bibinfo{author}{\bibfnamefont{N.~V.} \bibnamefont{Sarlis}},
  \bibinfo{author}{\bibfnamefont{E.~S.} \bibnamefont{Skordas}},
  \bibnamefont{and} \bibinfo{author}{\bibfnamefont{M.~S.}
  \bibnamefont{Lazaridou}}, \bibinfo{journal}{Phys. Rev. E}
  \textbf{\bibinfo{volume}{70}}, \bibinfo{pages}{011106}
  (\bibinfo{year}{2004}).

\bibitem[{\citenamefont{Varotsos
  et~al.}(2005{\natexlab{b}})\citenamefont{Varotsos, Sarlis, Skordas, and
  Lazaridou}}]{NAT05}
\bibinfo{author}{\bibfnamefont{P.~A.} \bibnamefont{Varotsos}},
  \bibinfo{author}{\bibfnamefont{N.~V.} \bibnamefont{Sarlis}},
  \bibinfo{author}{\bibfnamefont{E.~S.} \bibnamefont{Skordas}},
  \bibnamefont{and} \bibinfo{author}{\bibfnamefont{M.~S.}
  \bibnamefont{Lazaridou}}, \bibinfo{journal}{Phys. Rev. E}
  \textbf{\bibinfo{volume}{71}}, \bibinfo{pages}{011110}
  (\bibinfo{year}{2005}{\natexlab{b}}).

\bibitem[{\citenamefont{Varotsos
  et~al.}(2005{\natexlab{c}})\citenamefont{Varotsos, Sarlis, Tanaka, and
  Skordas}}]{NAT05B}
\bibinfo{author}{\bibfnamefont{P.~A.} \bibnamefont{Varotsos}},
  \bibinfo{author}{\bibfnamefont{N.~V.} \bibnamefont{Sarlis}},
  \bibinfo{author}{\bibfnamefont{H.~K.} \bibnamefont{Tanaka}},
  \bibnamefont{and} \bibinfo{author}{\bibfnamefont{E.~S.}
  \bibnamefont{Skordas}}, \bibinfo{journal}{Phys. Rev. E}
  \textbf{\bibinfo{volume}{71}}, \bibinfo{pages}{032102}
  (\bibinfo{year}{2005}{\natexlab{c}}).

\bibitem[{\citenamefont{Tanaka et~al.}(2004)\citenamefont{Tanaka, Varotsos,
  Sarlis, and Skordas}}]{TAN04}
\bibinfo{author}{\bibfnamefont{H.~K.} \bibnamefont{Tanaka}},
  \bibinfo{author}{\bibfnamefont{P.~V.} \bibnamefont{Varotsos}},
  \bibinfo{author}{\bibfnamefont{N.~V.} \bibnamefont{Sarlis}},
  \bibnamefont{and} \bibinfo{author}{\bibfnamefont{E.~S.}
  \bibnamefont{Skordas}}, \bibinfo{journal}{Proc. Japan Acad., Ser. B}
  \textbf{\bibinfo{volume}{80}}, \bibinfo{pages}{283} (\bibinfo{year}{2004}).

\bibitem[{\citenamefont{Lesche}(1982)}]{LES82}
\bibinfo{author}{\bibfnamefont{B.}~\bibnamefont{Lesche}}, \bibinfo{journal}{J.
  Stat. Phys.} \textbf{\bibinfo{volume}{27}}, \bibinfo{pages}{419}
  (\bibinfo{year}{1982}).

\bibitem[{\citenamefont{Lesche}(2004)}]{LES04}
\bibinfo{author}{\bibfnamefont{B.}~\bibnamefont{Lesche}},
  \bibinfo{journal}{Phys. Rev. E} \textbf{\bibinfo{volume}{70}},
  \bibinfo{pages}{017102} (\bibinfo{year}{2004}).

\bibitem[{\citenamefont{Bak et~al.}(2002)\citenamefont{Bak, Christensen, Danon,
  and Scanlon}}]{BAK02}
\bibinfo{author}{\bibfnamefont{P.}~\bibnamefont{Bak}},
  \bibinfo{author}{\bibfnamefont{K.}~\bibnamefont{Christensen}},
  \bibinfo{author}{\bibfnamefont{L.}~\bibnamefont{Danon}}, \bibnamefont{and}
  \bibinfo{author}{\bibfnamefont{T.}~\bibnamefont{Scanlon}},
  \bibinfo{journal}{Phys. Rev. Lett.} \textbf{\bibinfo{volume}{88}},
  \bibinfo{pages}{178501} (\bibinfo{year}{2002}).

\bibitem[{\citenamefont{Corral}(2004)}]{COR04}
\bibinfo{author}{\bibfnamefont{A.}~\bibnamefont{Corral}},
  \bibinfo{journal}{Phys. Rev. Lett.} \textbf{\bibinfo{volume}{92}},
  \bibinfo{pages}{108501} (\bibinfo{year}{2004}).

\bibitem[{\citenamefont{Baiesi and Paczuski}(2004)}]{BAI04}
\bibinfo{author}{\bibfnamefont{M.}~\bibnamefont{Baiesi}} \bibnamefont{and}
  \bibinfo{author}{\bibfnamefont{M.}~\bibnamefont{Paczuski}},
  \bibinfo{journal}{Phys. Rev. E} \textbf{\bibinfo{volume}{69}},
  \bibinfo{pages}{066106} (\bibinfo{year}{2004}).

\bibitem[{\citenamefont{Abe and Suzuki}(2004)}]{ABE04}
\bibinfo{author}{\bibfnamefont{S.}~\bibnamefont{Abe}} \bibnamefont{and}
  \bibinfo{author}{\bibfnamefont{N.}~\bibnamefont{Suzuki}},
  \bibinfo{journal}{Europhys. Lett.} \textbf{\bibinfo{volume}{65}},
  \bibinfo{pages}{581} (\bibinfo{year}{2004}).

\bibitem[{\citenamefont{Shebalin}(2006)}]{SHE06}
\bibinfo{author}{\bibfnamefont{P.}~\bibnamefont{Shebalin}},
  \bibinfo{journal}{Tectonophysics} \textbf{\bibinfo{volume}{424}},
  \bibinfo{pages}{335} (\bibinfo{year}{2006}).

\bibitem[{\citenamefont{Holliday et~al.}(2006)\citenamefont{Holliday, Rundle,
  Turcotte, Klein, Tiampo, and Donnellan}}]{HOL06}
\bibinfo{author}{\bibfnamefont{J.~R.} \bibnamefont{Holliday}},
  \bibinfo{author}{\bibfnamefont{J.~B.} \bibnamefont{Rundle}},
  \bibinfo{author}{\bibfnamefont{D.~L.} \bibnamefont{Turcotte}},
  \bibinfo{author}{\bibfnamefont{W.}~\bibnamefont{Klein}},
  \bibinfo{author}{\bibfnamefont{K.~F.} \bibnamefont{Tiampo}},
  \bibnamefont{and}
  \bibinfo{author}{\bibfnamefont{A.}~\bibnamefont{Donnellan}},
  \bibinfo{journal}{Phys. Rev. Lett.} \textbf{\bibinfo{volume}{97}},
  \bibinfo{pages}{238501} (\bibinfo{year}{2006}).

\bibitem[{\citenamefont{Tiampo et~al.}(2007)\citenamefont{Tiampo, Rundle,
  Klein, Holliday, Martins, and Ferguson}}]{TIA07}
\bibinfo{author}{\bibfnamefont{K.~F.} \bibnamefont{Tiampo}},
  \bibinfo{author}{\bibfnamefont{J.~B.} \bibnamefont{Rundle}},
  \bibinfo{author}{\bibfnamefont{W.}~\bibnamefont{Klein}},
  \bibinfo{author}{\bibfnamefont{J.}~\bibnamefont{Holliday}},
  \bibinfo{author}{\bibfnamefont{J.~S.~S.} \bibnamefont{Martins}},
  \bibnamefont{and} \bibinfo{author}{\bibfnamefont{C.~D.}
  \bibnamefont{Ferguson}}, \bibinfo{journal}{Phys. Rev. E}
  \textbf{\bibinfo{volume}{75}}, \bibinfo{pages}{066107}
  (\bibinfo{year}{2007}).

\bibitem[{\citenamefont{Klein et~al.}(2007)\citenamefont{Klein, Gould,
  Gulbahce, Rundle, and Tiampo}}]{KLE07}
\bibinfo{author}{\bibfnamefont{W.}~\bibnamefont{Klein}},
  \bibinfo{author}{\bibfnamefont{H.}~\bibnamefont{Gould}},
  \bibinfo{author}{\bibfnamefont{N.}~\bibnamefont{Gulbahce}},
  \bibinfo{author}{\bibfnamefont{J.~B.} \bibnamefont{Rundle}},
  \bibnamefont{and} \bibinfo{author}{\bibfnamefont{K.}~\bibnamefont{Tiampo}},
  \bibinfo{journal}{Phys. Rev. E} \textbf{\bibinfo{volume}{75}},
  \bibinfo{pages}{031114} (\bibinfo{year}{2007}).

\bibitem[{\citenamefont{Sornette}(2004)}]{SOR04}
\bibinfo{author}{\bibfnamefont{D.}~\bibnamefont{Sornette}},
  \emph{\bibinfo{title}{Critical Phenomena in Natural Science}}
  (\bibinfo{publisher}{Springer}, \bibinfo{address}{Berlin},
  \bibinfo{year}{2004}), \bibinfo{edition}{2nd} ed.

\bibitem[{\citenamefont{Uyeda et~al.}(1999)\citenamefont{Uyeda, Al-Damegh,
  Dologlou, and Nagao}}]{DOL99}
\bibinfo{author}{\bibfnamefont{S.}~\bibnamefont{Uyeda}},
  \bibinfo{author}{\bibfnamefont{K.~S.} \bibnamefont{Al-Damegh}},
  \bibinfo{author}{\bibfnamefont{E.}~\bibnamefont{Dologlou}}, \bibnamefont{and}
  \bibinfo{author}{\bibfnamefont{T.}~\bibnamefont{Nagao}},
  \bibinfo{journal}{Tectonophysics} \textbf{\bibinfo{volume}{304}},
  \bibinfo{pages}{41} (\bibinfo{year}{1999}).

\bibitem[{\citenamefont{Varotsos
  et~al.}(2005{\natexlab{d}})\citenamefont{Varotsos, Sarlis, Skordas, and
  Lazaridou}}]{TEC05}
\bibinfo{author}{\bibfnamefont{P.}~\bibnamefont{Varotsos}},
  \bibinfo{author}{\bibfnamefont{N.}~\bibnamefont{Sarlis}},
  \bibinfo{author}{\bibfnamefont{E.}~\bibnamefont{Skordas}}, \bibnamefont{and}
  \bibinfo{author}{\bibfnamefont{M.}~\bibnamefont{Lazaridou}},
  \bibinfo{journal}{Tectonophysics} \textbf{\bibinfo{volume}{412}},
  \bibinfo{pages}{279} (\bibinfo{year}{2005}{\natexlab{d}}).

\bibitem[{\citenamefont{Varotsos et~al.}(2007)\citenamefont{Varotsos, Sarlis,
  and Skordas}}]{vers1}
\bibinfo{author}{\bibfnamefont{P.~A.} \bibnamefont{Varotsos}},
  \bibinfo{author}{\bibfnamefont{N.~V.} \bibnamefont{Sarlis}},
  \bibnamefont{and} \bibinfo{author}{\bibfnamefont{E.~S.}
  \bibnamefont{Skordas}} (\bibinfo{year}{2007}), \eprint{arXiv:0711.3766v1}.

\bibitem[{\citenamefont{Varotsos
  et~al.}(2008{\natexlab{b}})\citenamefont{Varotsos, Sarlis, and
  Skordas}}]{vers3}
\bibinfo{author}{\bibfnamefont{P.~A.} \bibnamefont{Varotsos}},
  \bibinfo{author}{\bibfnamefont{N.~V.} \bibnamefont{Sarlis}},
  \bibnamefont{and} \bibinfo{author}{\bibfnamefont{E.~S.}
  \bibnamefont{Skordas}} (\bibinfo{year}{2008}{\natexlab{b}}),
  \eprint{arXiv:0711.3766v3}.

\bibitem[{\citenamefont{Sarlis et~al.}(2008)\citenamefont{Sarlis, Skordas,
  Lazaridou, and Varotsos}}]{SAR08A}
\bibinfo{author}{\bibfnamefont{N.~V.} \bibnamefont{Sarlis}},
  \bibinfo{author}{\bibfnamefont{E.~S.} \bibnamefont{Skordas}},
  \bibinfo{author}{\bibfnamefont{M.~S.} \bibnamefont{Lazaridou}},
  \bibnamefont{and} \bibinfo{author}{\bibfnamefont{P.~A.}
  \bibnamefont{Varotsos}} (\bibinfo{year}{2008}), \eprint{arXiv:0802.3329v1}.

\end{thebibliography}
\bibliographystyle{apsrev}

\begin{turnpage}
\begin{figure*}
\includegraphics{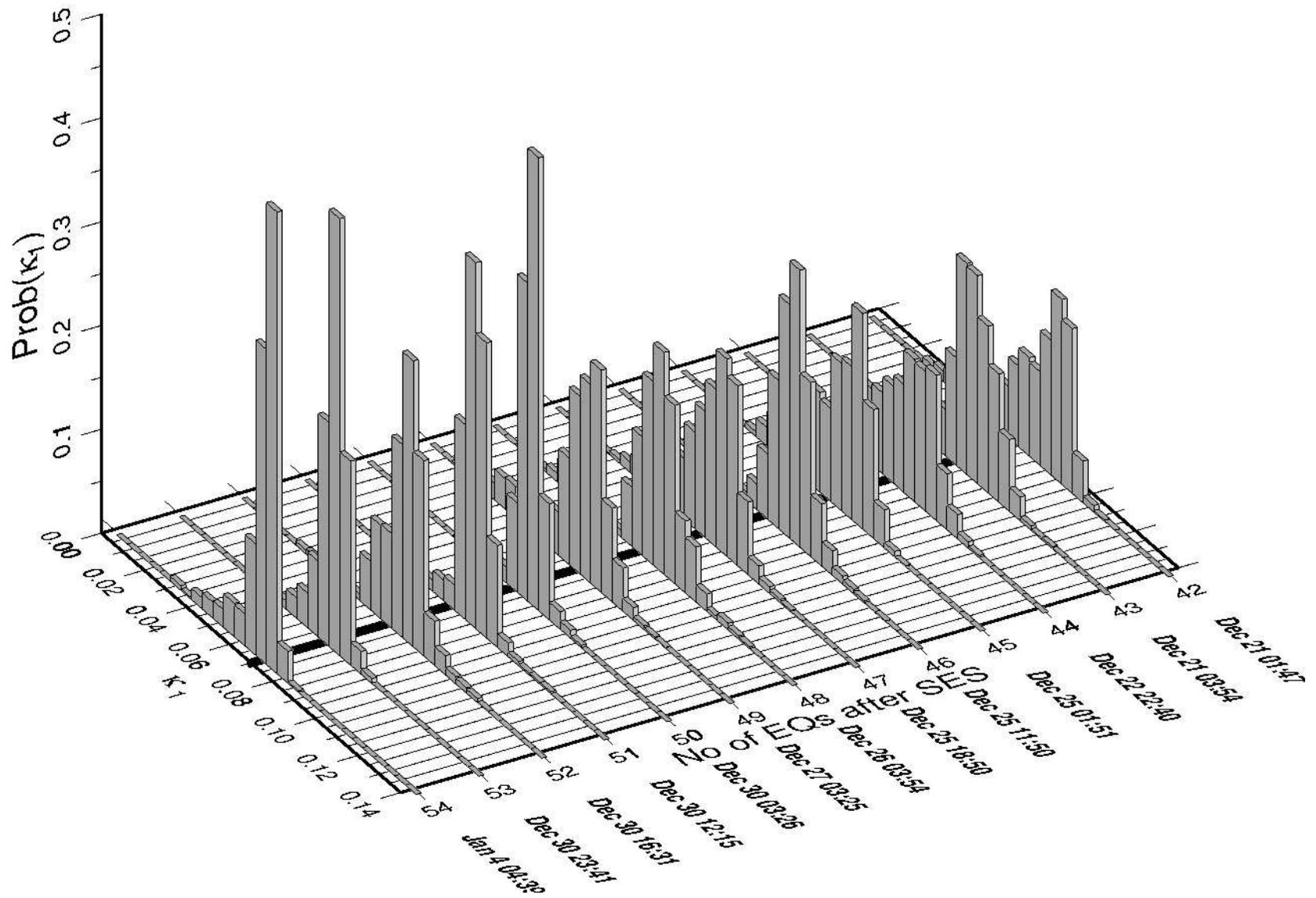}
\caption{Study of the Prob($\kappa_1$) for the seismicity that occurred within the area $N_{37.6}^{38.6}E_{20.0}^{23.3}$ after the SES activity at PAT on November 7, 2007.} \label{Leo}
\end{figure*}

\begin{figure*}
\includegraphics{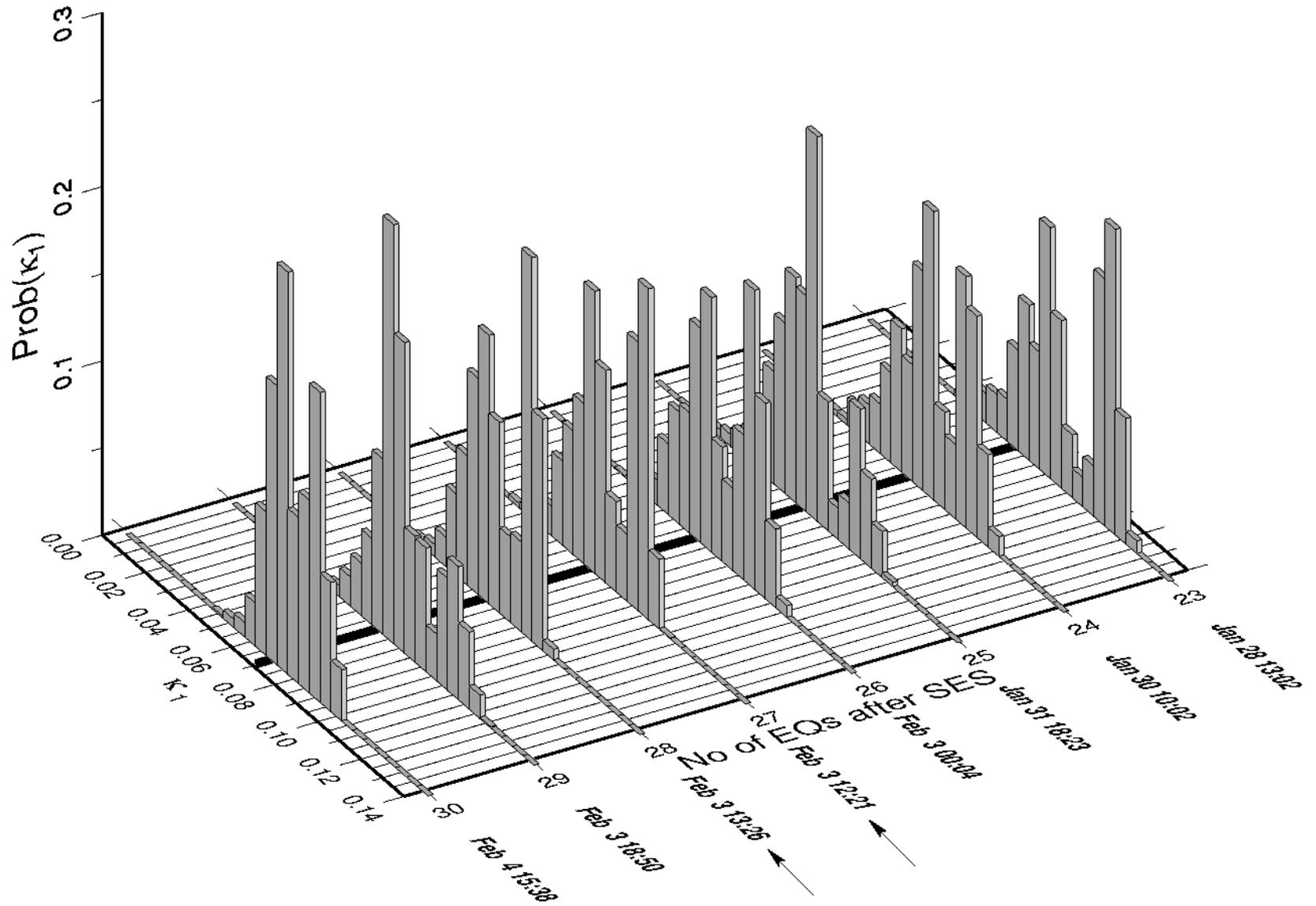}
\caption{The same as Fig.\ref{Leo}, but for the SES activity at PAT on January 10, 2008.} \label{Pat}
\end{figure*}

\begin{figure*}
\includegraphics{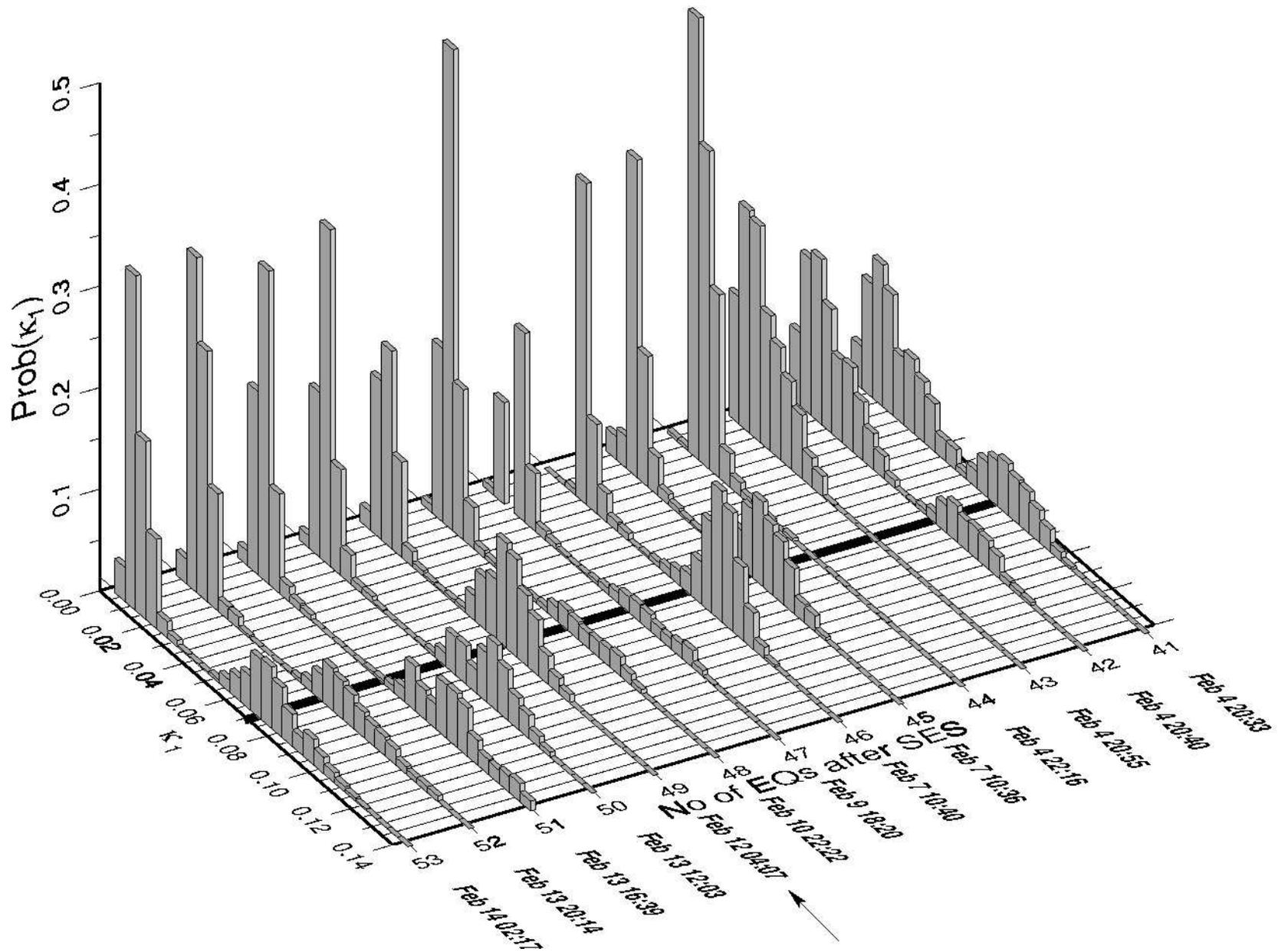}
\caption{The same as Fig.\ref{Leo}, but for the area $N_{36.0}^{38.6}E_{20.0}^{22.5}$ after the SES activity at PIR on January 14, 2008.} \label{p14}
\end{figure*}

\begin{figure*}
\includegraphics{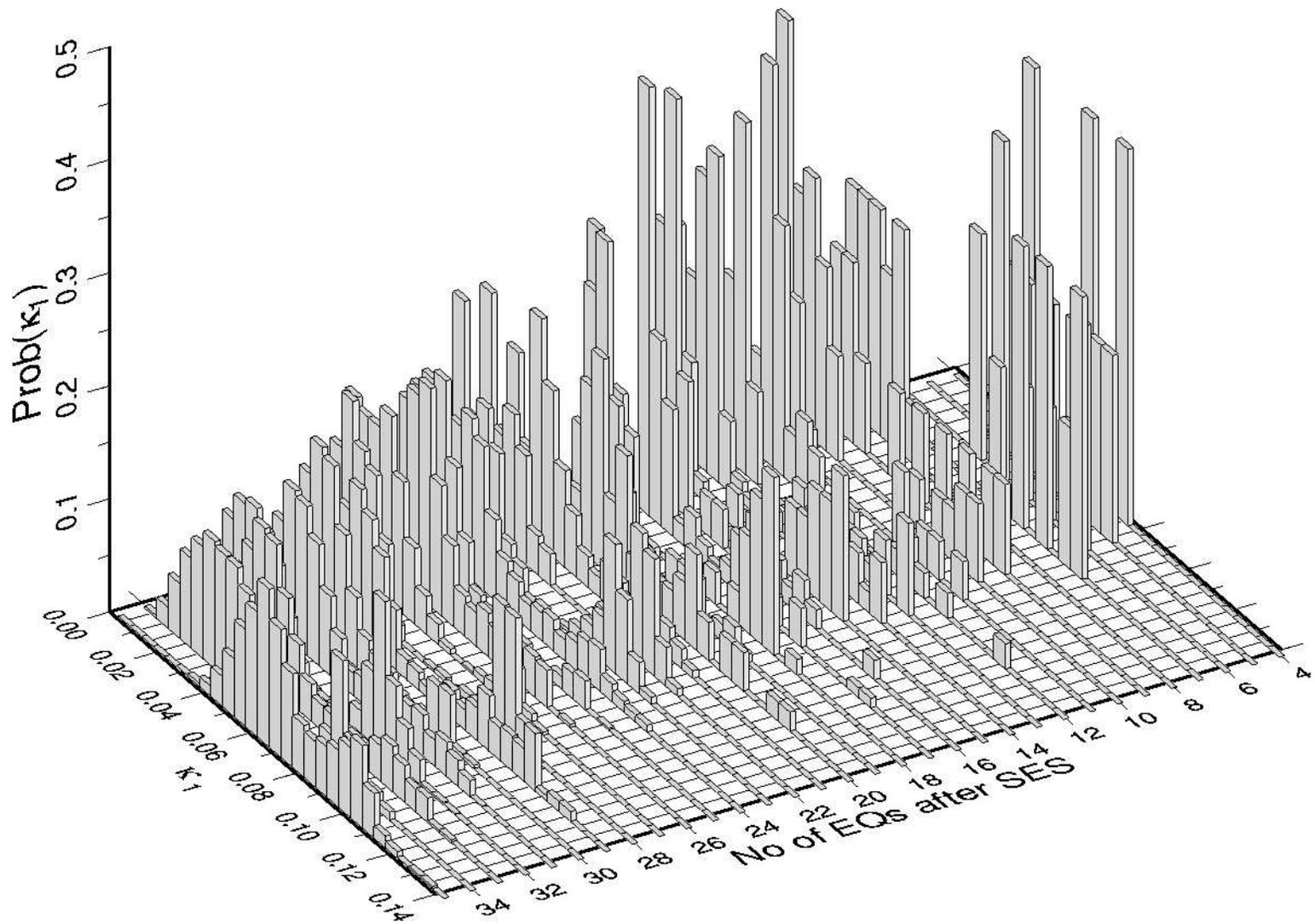}
\caption{The same as Fig.\ref{Leo}, but for the area $N_{37.5}^{38.6}E_{20.0}^{23.3}$ after the SES activity at PAT on February 9, 2008.} \label{oldfig5}
\end{figure*}

\begin{figure*}
\includegraphics{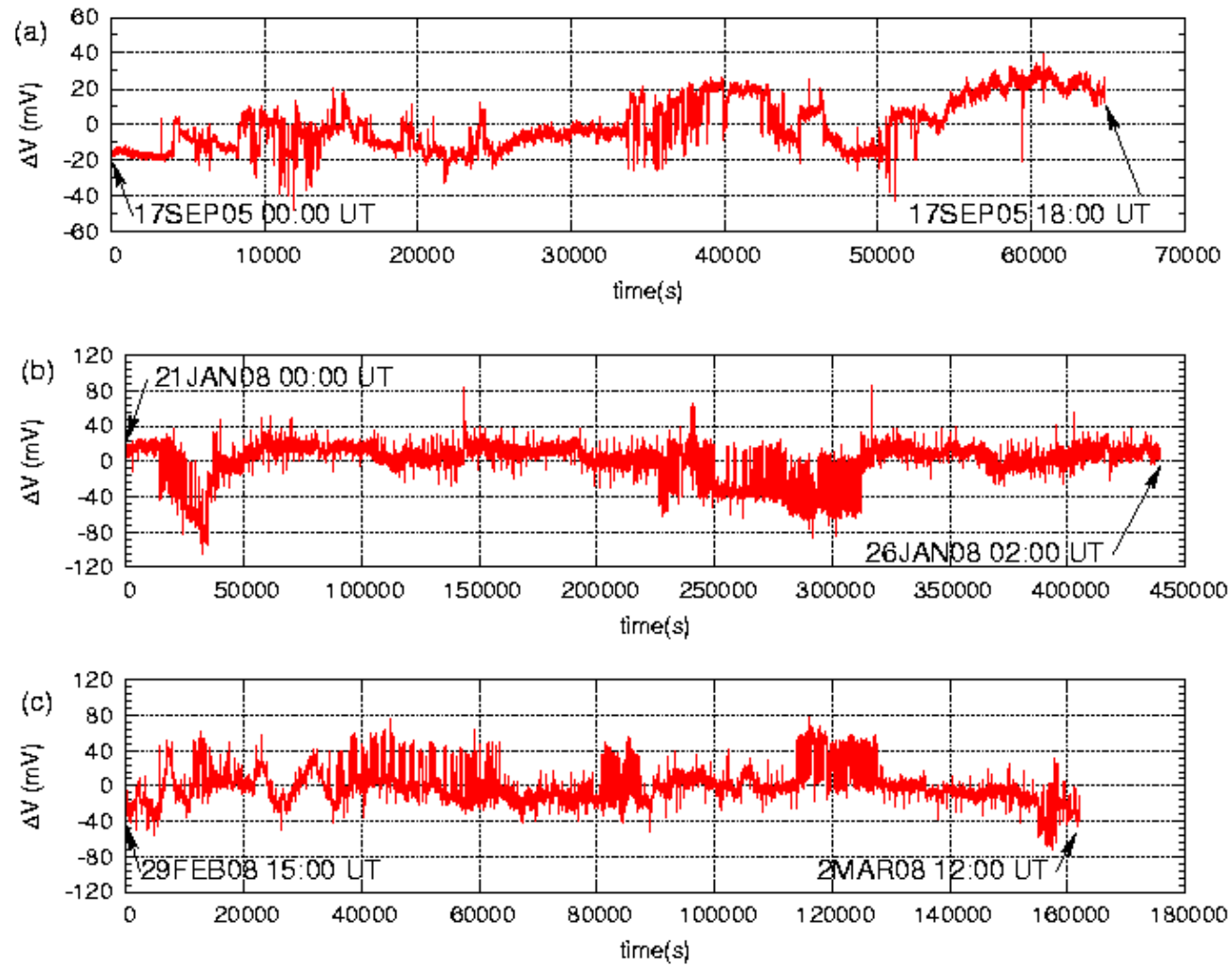}
\caption{The most recent SES activity recorded at PIR (c) along with the one that preceded the $M_w 6.9$ EQ on February 14, 2008 (b) and the $M_w 6.7$ EQ on January 8, 2006 (a).} \label{fig5}
\end{figure*}

\begin{figure*}
\includegraphics{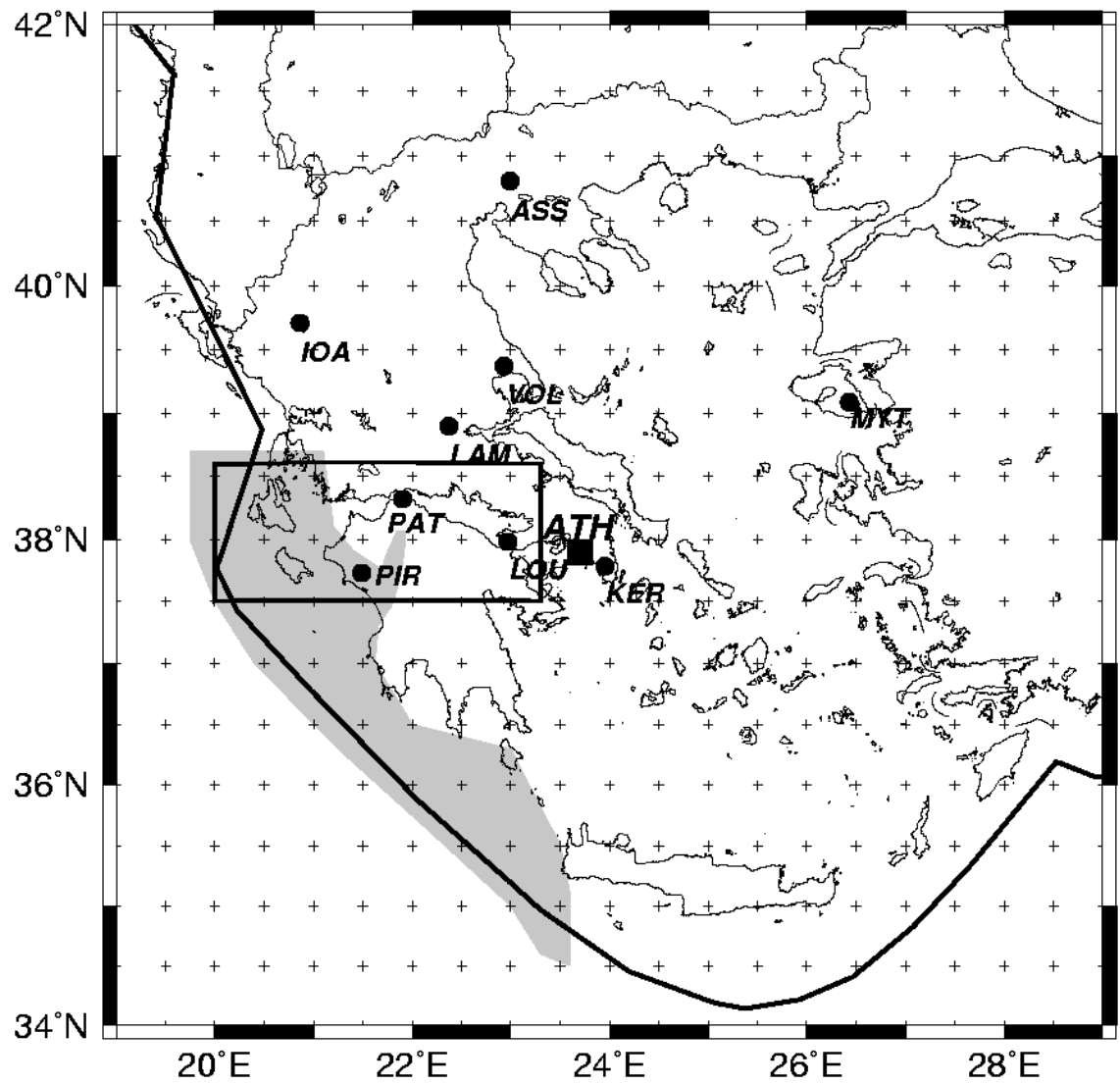}
\caption{The shaded area shows the up to date knowledge of the PIR selectivity map.}
\label{fig6}
\end{figure*}

\begin{figure*}
\includegraphics{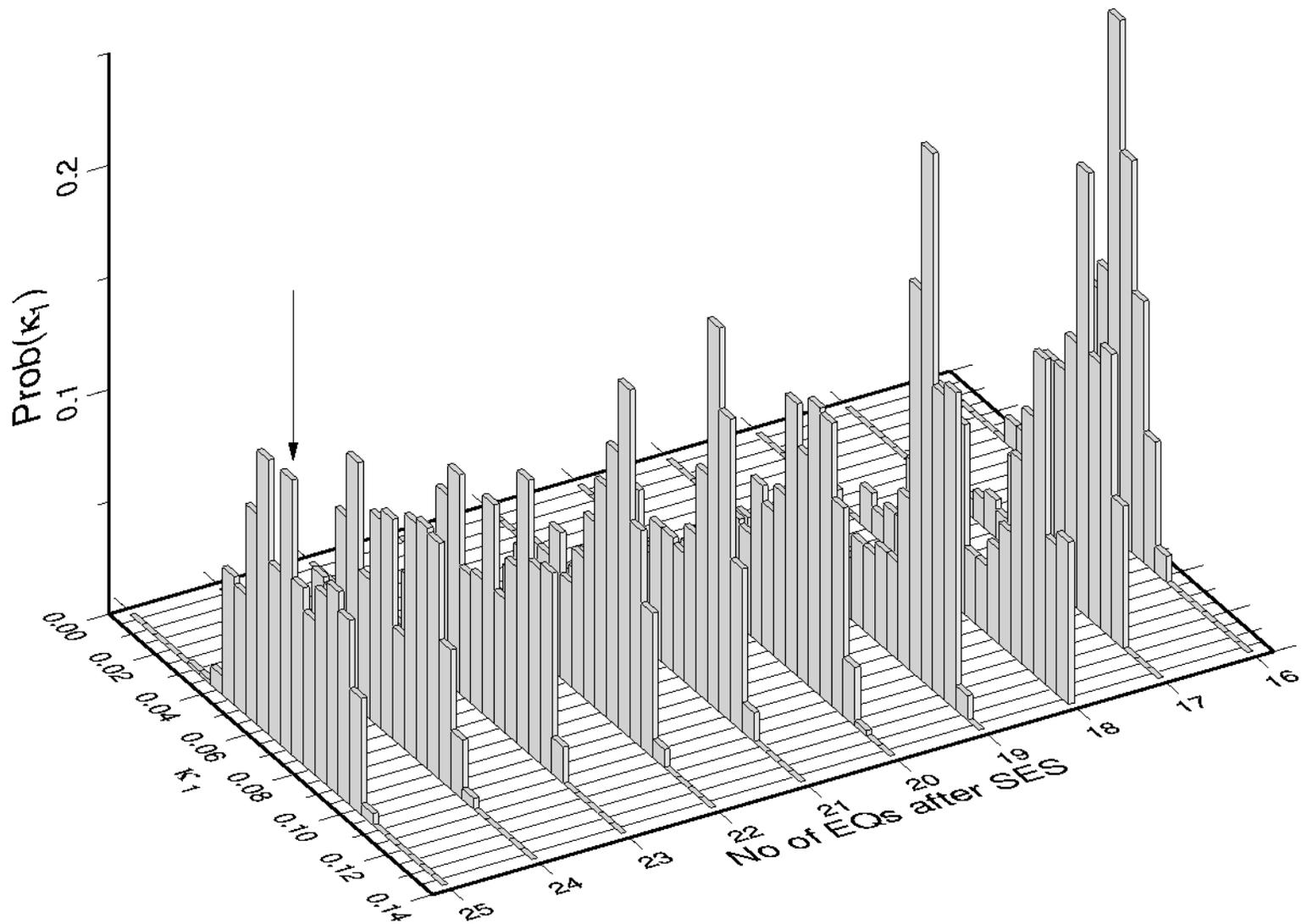}
\caption{The arrow shows that the Prob($\kappa_1$) of the seismicity (subsequent to the long duration SES activity recorded at PIR during February 29 to March 2, 2008) within the area 
$N_{37.0}^{38.6}E_{20.0}^{22.0}$ maximized at $\kappa_1\approx$0.070 upon the occurrence of a small M3.2 event at 19:16 UT on March 25, 2008(i.e.,on the occurence of the 25th event after the SES).}
\label{figa1}
\end{figure*}

\begin{figure*}
\includegraphics{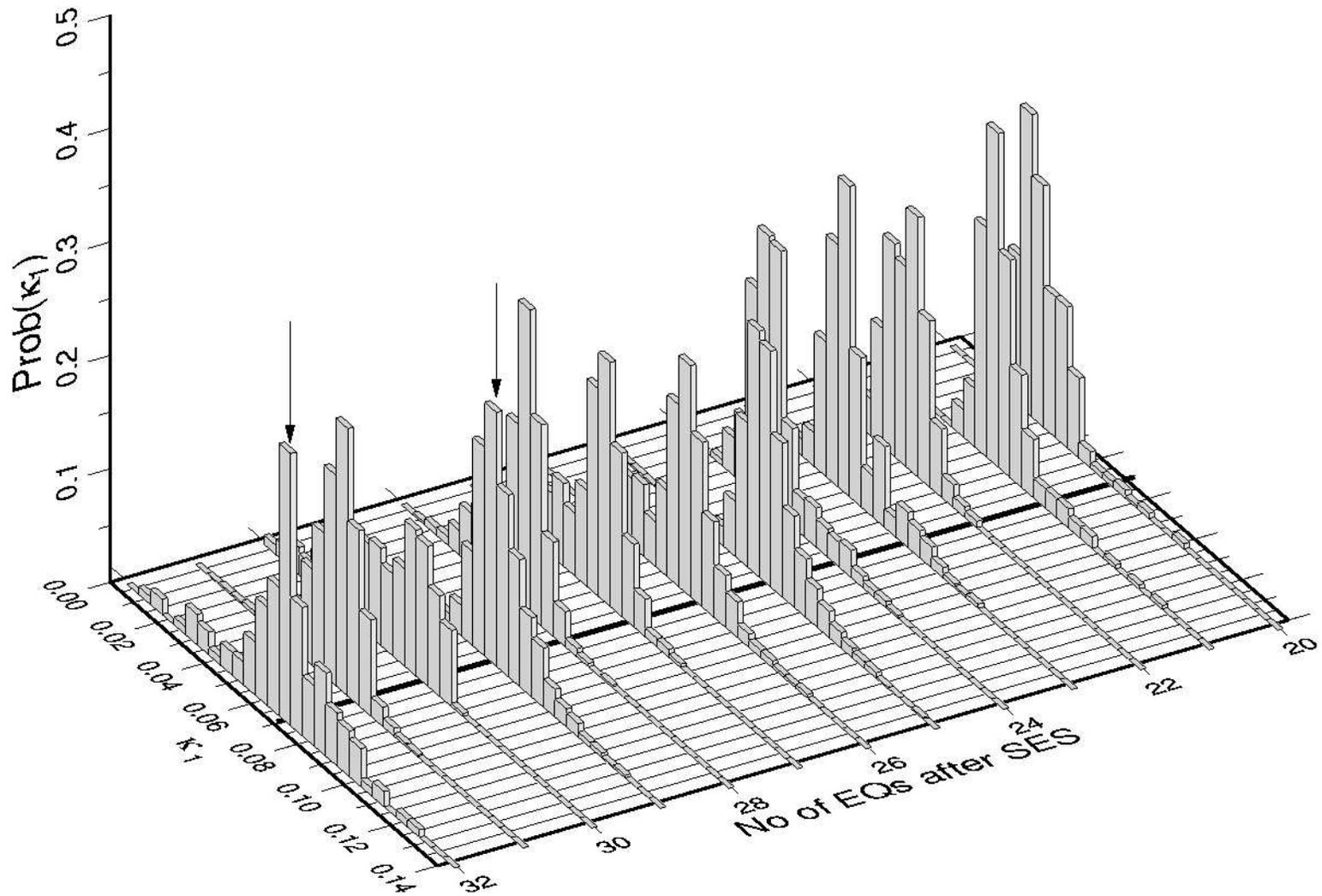}
\caption{Prob($\kappa_1$) versus $\kappa_1$ of the seismicity, for $M_{thres}=3.9$, (subsequent to the long duration SES activity recorded at PIR during February 29 to March 2, 2008) within the shaded area shown in Fig.\ref{fig6}. The two arrows mark the maxima at $\kappa_1=0.07$ that occurred on May 8, 2008 (i.e., on the occurence of the 29th event after the SES) and on May 27, 2008 (i.e.,on the occurence of the 32nd event after the SES). The first maximum 
has been followed by the 5.6EQ on May 10, 2008, as described in the Appendix.} \label{figa2}
\end{figure*}

\begin{figure*}
\includegraphics{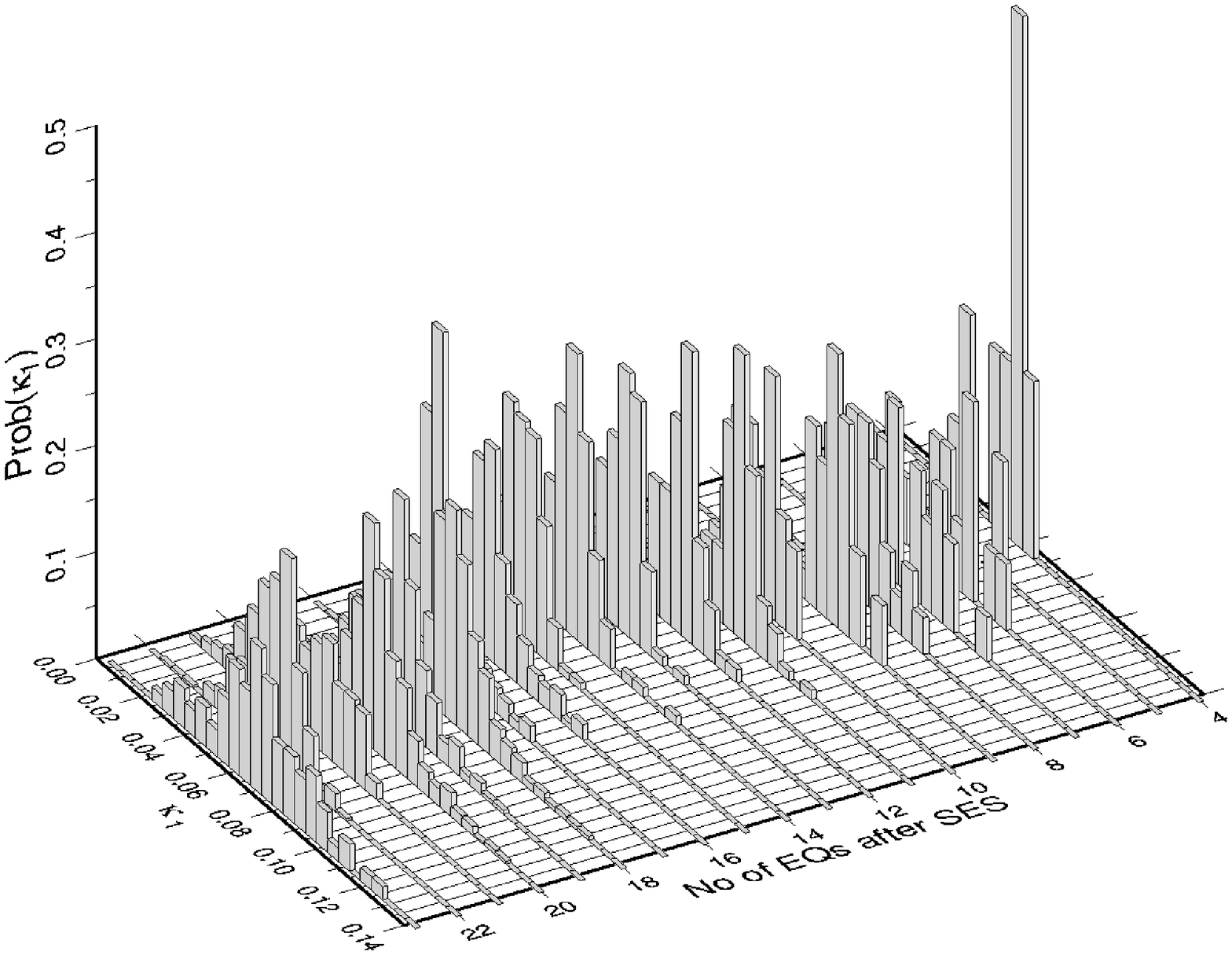}
\caption{The same as Fig.\ref{figa2}, but for $M_{thres}=4.0$. The last histogram corresponds to the 5.1 event on May 27, 2008 and exhibits a maximum at $\kappa_1=0.07$.} \label{figa3}
\end{figure*}

\begin{figure*}
\includegraphics{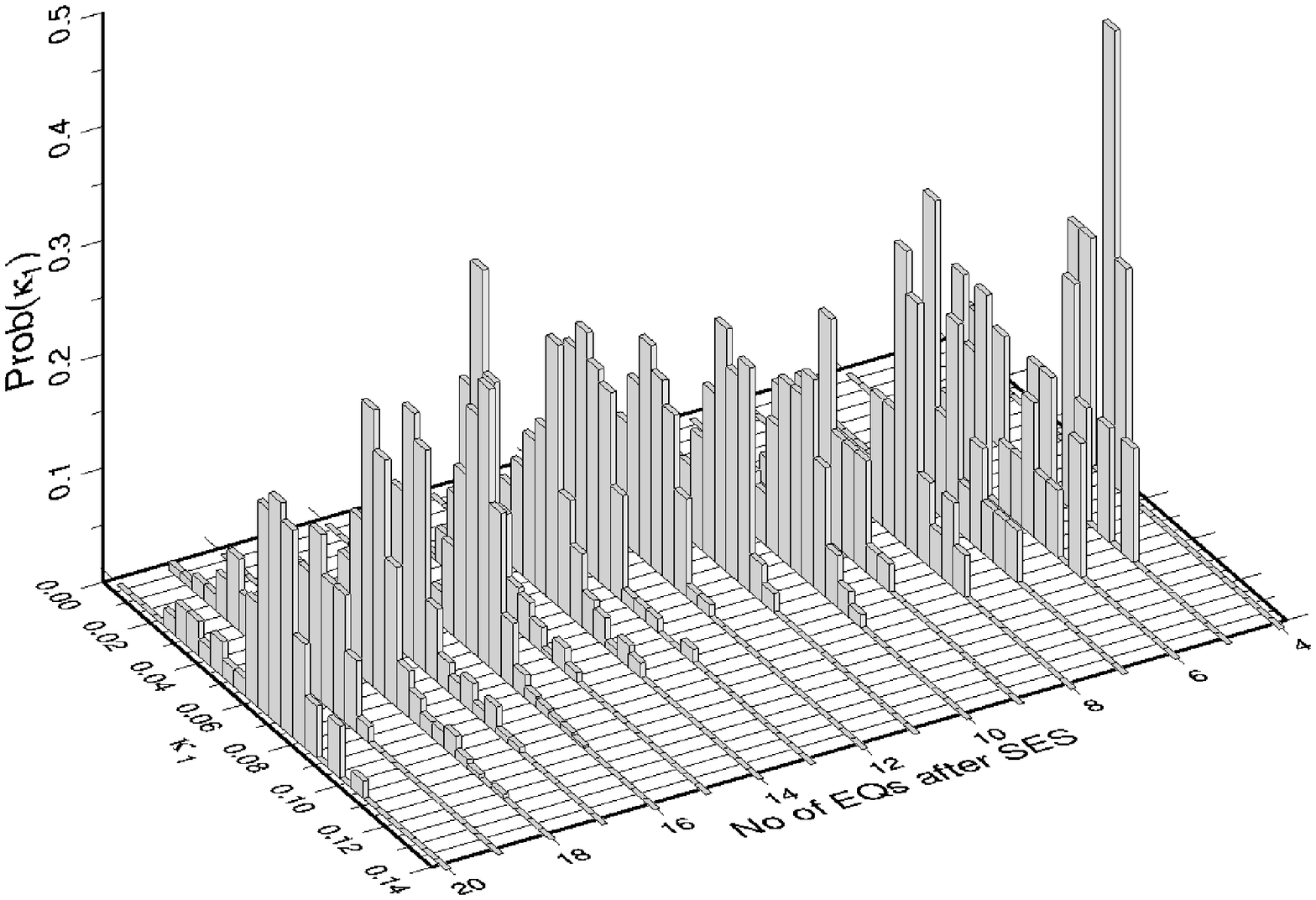}
\caption{The same as Fig.\ref{figa2}, but for $M_{thres}=4.1$. The last histogram corresponds to the 5.1 event on May 27, 2008 and exhibits a maximum at $\kappa_1=0.07$.} \label{figa4}
\end{figure*}

\end{turnpage}

\end{document}